%% file: real_space_RPA.tex
\documentclass[preprint, superscriptaddress,amsmath,amssymb,aps,prb]{revtex4-2}

\usepackage{siunitx}
\usepackage{makecell}
\usepackage[colorlinks,linkcolor=black,anchorcolor=blue, urlcolor=blue,citecolor=blue,unicode]{hyperref}
\usepackage{multirow}
\usepackage{subfigure}
\usepackage{color}
\usepackage[normalem]{ulem}
\usepackage{amsmath}
\usepackage{booktabs,siunitx,array}
\usepackage{booktabs}
\usepackage{ulem}
\usepackage{listings}
\usepackage{import}
\usepackage{xcolor}
\usepackage{framed}
\usepackage{mdwlist}
\usepackage{diagbox}
\usepackage{bm}
\usepackage{graphicx}
\usepackage{dcolumn}
\usepackage{algorithm}
\usepackage{algpseudocode}
\usepackage{csquotes}
\usepackage{array}
\usepackage{subfigure}
\usepackage{siunitx}
\usepackage{amsmath}
\usepackage{ifthen}
\usepackage{threeparttable}
\usepackage{float}

\usepackage{algorithmicx}
\usepackage{booktabs}
\usepackage{cleveref}
\usepackage{blindtext}

\usepackage{enumerate}

\input{newcommands}

\begin{document}

\title{Sub-quadratic scaling real-space random-phase approximation correlation energy calculations for periodic systems with numerical atomic orbitals}

\author{Rong Shi}
\affiliation{CAS Key Laboratory of Quantum Information, University of Science and Technology of
China, Hefei 230026, Anhui, China}
\affiliation{Institute of Physics, Chinese Academy of Sciences, Beijing 100190, China}

\author{Peize Lin}
\affiliation{Institute of Physics, Chinese Academy of Sciences, Beijing 100190, China}
\affiliation{Songshan Lake Materials Laboratory, Dongguan 523808, Guangdong, China}

\author{Min-Ye Zhang}
\affiliation{Institute of Physics, Chinese Academy of Sciences, Beijing 100190, China}

\author{Lixin He}
\email{helx@ustc.edu.cn}
\affiliation{CAS Key Laboratory of Quantum Information, University of Science and Technology of
China, Hefei 230026, Anhui, China}

\author{Xinguo Ren}
\email{renxg@iphy.ac.cn}
\affiliation{Institute of Physics, Chinese Academy of Sciences, Beijing 100190, China}
\affiliation{Songshan Lake Materials Laboratory, Dongguan 523808, Guangdong, China}

\begin{abstract}
The random phase approximation (RPA) as formulated as an orbital-dependent, fifth-rung functional within the density functional theory (DFT) framework offers a promising approach for calculating the ground-state energies and the derived properties of real materials. Its widespread use to large-size, complex materials is
however impeded by the significantly increased computational cost, compared to lower-rung functionals. The standard implementation exhibits an $\mathcal{O}(N^4)$-scaling behavior with respect to system size $N$. In this work,
we develop a low-scaling RPA algorithm for periodic systems, based on
the numerical atomic orbital (NAO) basis-set framework and a localized variant of the resolution of identity (RI) approximation. 
The rate-determining step for RPA calculations -- the evaluation of non-interacting response function matrix, is reduced from
$\mathcal{O}(N^4)$ to $\mathcal{O}(N^2)$ by just exploiting the sparsity of the RI expansion coefficients, resultant from 
localized RI (LRI) scheme and the strict locality of NAOs.
The computational cost of this step can be further reduced to linear scaling if the decay behavior of the Green's function 
in real space can be further taken into account. 
Benchmark calculations against existing $\bfk$-space based implementation confirms the validity and high numerical precision
of the present algorithm and implementation. The new RPA algorithm allows us to readily handle three-dimensional, close-packed
solid state materials with
over 1000 atoms. The algorithm and numerical techniques developed in this work also have implications for developing low-scaling algorithms
for other correlated methods to be applicable to large-scale extended materials.


\end{abstract}

\maketitle 

\section{Introduction}

Random phase approximation (RPA) \cite{Bohm/Pines:1953,Gell-Mann/Brueckner:1957,Hubbard:1957b} as formulated within the framework of adiabatic-correction fluctuation-dissipation theorem (ACFDT) \cite{Langreth/Perdew:1977,Gunnarsson/Lundqvist:1976} provides 
an appealing approach to compute the ground-state energy of interacting many-electron systems \cite{Furche:2001,Fuchs/Gonze:2002,Eshuis/Bates/Furche:2012,Ren/etal:2012b}. It can be viewed as a non-local approximation
for the exchange-correlation (XC) energy functional within Kohn-Sham (KS) density functional theory (DFT) \cite{Hohenberg/Kohn:1964,Kohn/Sham:1965}. According to the Jacob's ladder 
classifying different XC functionals \cite{Perdew/Schmidt:2001}, the RPA sits on the top rung of the ladder, and captures seamlessly non-local 
many-electron correlations that are missing in lower-rung functionals. Applications of RPA to real materials show that this approach 
performs rather well
in describing energy differences, in particular the surface adsorption energies \cite{Ren/etal:2009,Schimka/etal:2010}, the reaction barrier heights \cite{Paier/etal:2012,Ren/etal:2013}, and the delicate energy differences between different
polymorphs \cite{Lebegue/etal:2010,sun:2019,Sengupta/etal:2018,Cazorla/Gould:2019,Yang/Ren:2022}. Despite its promising performance, a widespread use of RPA is hindered by its quickly increasing computational cost with system size.
To deal with this issue, a considerable amount of recent works are devoted to developing low-scaling
algorithms to speed up the RPA calculations,
\cite{Neuhauser/etal:2013,Moussa:2014,Kaltak/Klimes/Kresse:2014,Kallay:2015,Wilhelm/etal:2016,Graf/etal:2018,Luenser/Schurkus/Ochsenfeld:2017,Lu/Thicke:2017,Duchemin/Blase:2019}, paving ways for applying RPA to large-scale, complex materials that 
are previously out of reach.

The key quantity in RPA calculations is the non-interacting KS density response function $\chi^0$, represented within a suitable basis set. The standard computational scaling for evaluating $\chi^0$ is $\mathcal{O}(N^4)$ with $N$ being a measure of system size, for both plane-wave basis sets and
the resolution-of-identity (RI) formulation of RPA within atomic-orbital basis sets.  The $\mathcal{O}(N^4)$ scaling can be reduced to $\mathcal{O}(N^3)$ by utilizing
the space-time algorithm \cite{Rojas/Godby/Needs:1995,White/Godby/Rieger/Needs:1997,Rieger/etal:1999}, initially developed for the $GW$ method 
\cite{Hedin:1965}. Thanks to the development of the minimax
quadrature grid by Kaltak \textit {et al.} \cite{Kaltak/Klimes/Kresse:2014,Kaltak/etal:2014b} which enables an efficient discrete Fourier transform from the imaginary time domain to the imaginary frequency domain, the $\mathcal{O}(N^3)$ algorithm becomes superior to the standard $\mathcal{O}(N^4)$ one at a 
cross point of system size that can be handled by modern computers. Such a dual real-space and plane-wave formulation of $\mathcal{O}(N^3)$ RPA 
(and analogously $GW$ \cite{PeitaoLiu/etal:2016}) algorithm
was soon extended to Gaussian atomic-orbital framework, combined with the RI technique of different flavors, 
like the overlap-metric \cite{Wilhelm/etal:2016} and attenuated-Coulomb-metric \cite{Wilhelm/Seewald/Golze:2021} based RI schemes, 
pair atomic density fitting \cite{Forster/Visscher:2020}, and the interpolative separable RI scheme 
\cite{Lu/Yin:2015,Lu/Thicke:2017,Duchemin/Blase:2019}. Benefited further from the spatial locality of atomic orbitals, 
algorithms and implementations
with $\mathcal{O}(N)$ to $\mathcal{O}(N^3)$ scaling behaviors have been reported \cite{Wilhelm/etal:2016,Graf/etal:2018,Luenser/Schurkus/Ochsenfeld:2017,Lu/Thicke:2017,Duchemin/Blase:2019}. Apart from these, 
radically different approaches based on solving the Riccati equation using the local correlation method \cite{Kallay:2015}, as well as 
on a stochastic formulation of ACFDT-RPA via time-dependent DFT \cite{Neuhauser/etal:2013,Gao/Neuhauer/etal:2015} have been developed, 
allowing for linear or even sublinear RPA correlation calculations. Practically, the plane-wave based implementations are more suitable for 
describing periodic systems, whereas the atomic-orbital based implementations are typically applied to finite molecular systems and/or supercell-based $\Gamma$-only simulations. 

In this work, we present yet another low-scaling algorithm for periodic RPA calculations  with finite $\bfk$-point sampling
using numerical atomic orbital (NAO) basis sets. 
In this algorithm, the computational cost for the key step of the RPA calculations, namely, the evaluation of the KS response function matrix, 
scales quadratically or better with respect to the number of atoms in the unit cell and linearly with the number of $\bfk$ points 
in the Brillouin zone. This is enabled by the localized resolution of identity (LRI) approximation \cite{Ihrig/etal:2015}, 
a prescreening of the sparse RI coefficients \cite{Levchenko/etal:2015,Lin/Ren/He:2020,Lin/Ren/He:2021},
and an efficient imaginary time-to-frequency Fourier transform using the minimax grid \cite{Kaltak/Klimes/Kresse:2014,Kaltak/etal:2014b}. 
The $\mathcal{O}(N^2)$ scaling can in fact be made asymptotically linear
for insulating systems, if the spatial decay of the Green's function is further taken into account. The algorithm has been implemented in a standalone
library package called LibRPA, which has been interfaced with two NAO-based first-principles codes FHI-aims \cite{Blum/etal:2009} 
and ABACUS \cite{Chen/Guo/He:2010,Li/Liu/etal:2016,abacusweb}. The development of LibRPA allows one to do efficient
RPA calculations with NAO-based first-principles codes, with necessary inputs provided by the latter.

The LRI approximation, which is crucial for the design of low-scaling algorithms, has been used in periodic hybrid functional \cite{Levchenko/etal:2015,Lin/Ren/He:2020,Lin/Ren/He:2021} and $G_0W_0$ 
calculations \cite{Ren/etal:2021}, as well as in RPA force calculations for molecules \cite{Tahir/etal:2022} before. Various benchmark
calculations showed that this approximation can be made sufficiently accurate, provided that high-quality auxiliary basis sets can be
constructed \cite{Ihrig/etal:2015,Levchenko/etal:2015,Lin/Ren/He:2020,Lin/Ren/He:2021,Tahir/etal:2022}. While the accuracy of LRI for 
NAO-based periodic RPA calculations will be benchmarked elsewhere, here we mainly focus on the low-scaling algorithm, the implementation
details, and the scaling behavior with respect to both system size and the number of $\bfk$ points. Thanks to the existing canonical $\mathcal{O}(N^4)$-scaling periodic
RPA implementation in FHI-aims, the accuracy and efficacy of the low-scaling RPA implementation can be unambiguously benchmarked. We show that
our present implementation can readily treat 3-dimensional bulk systems containing over 1000 atoms, with reasonable computational resources. 

The paper is organized as follows. The key equations behind the NAO-based low-scaling RPA algorithm are presented in Sec.~\ref{sec:theory},
which is followed by Sec.~\ref{sec:implementation} which contains a detailed discussion of the actual loop structure adopted 
in the low-scaling algorithm and the implementation details.
Section~\ref{sec:results} presents the major results, consisting of test calculations that validate the algorithm and implementation by
comparing to the existing $\bfk$-space based algorithm in FHI-aims, and benchmarks of the scaling behavior of 
the computational cost with respect to system size. In addition, we also discuss the importance of 
incorporating the sparsity of the Green's function in the algorithm, which brings significant further reduction of the computational cost. 
Finally, we report a scaling-behavior study for system sizes beyond 1000 atoms by interfacing LibRPA with 
another NAO-based DFT code -- ABACUS. The Appendix presents a detailed derivation of the key equations in Sec.~\ref{sec:theory}, and
the decay behavior of the Green's function in real space 
for prototypical systems.

\section{Theoretical formulation}
\label{sec:theory}

In this section, we will present the key equations behind the low-scaling algorithm of periodic RPA within the NAO basis-set framework.
The formalism should be applicable to Gaussian-type or other types of localized atomic orbitals as well, provided that high-quality auxiliary basis sets (ABSs) are available and the LRI is sufficiently accurate.

Within the ACFDT framework, the RPA correlation energy is formally given by
\cite{RN171} 
\begin{equation}
    \label{eq:cRPA}
    E^\textrm{RPA}_c=\frac
    {1}{2\pi}\int^{\infty}_0 \textrm{d}\omega\, \textrm{Tr}\left[\textrm{ln}(1-\chi^0(\ii\omega)v)+\chi^0(\ii\omega)v\right]\, ,
\end{equation}
where $\chi^0$ represents the KS independent density response function on the imaginary frequency axis and $v$ the bare Coulomb potential.
For a periodic system, the spatially non-local function $\chi^0(\bfr,\bfrp,\ii\omega)$ can be represented in terms of a set of Bloch-summed
atom-centered auxiliary basis functions (ABFs),
\begin{equation}
    \chi^0(\bfr,\bfrp,\ii\omega) = \frac{1}{N_\bfk}\sum_{\mu,\nu,\bfq}P_{\mu}^\bfq(\bfr) \chi^0_{\mu\nu}(\bfq,\ii\omega) P_{\nu}^{\bfq\ast}(\bfrp)
    \label{eq:chi0_k_aux-expan}
\end{equation}
where the summation over $\bfq$ goes over the first Brillouin zone (BZ), and $N_\bfk$ is the number of $\bfk$ points in the 1st BZ.
In Eq.~\eqref{eq:chi0_k_aux-expan}, 
\begin{equation}
P_{\mu}^\bfq(\bfr)=\sum_{\bfR} \ee^{\ii\bfq\cdot\bfR} P(\bfr-{\bm\tau}_{\cal U}-\bfR) \, ,
     \label{eq:aux_k-space}
\end{equation}
with ${\cal U}$ denoting the atom on which the ABF $P_\mu(\bfr)$ is sitting and ${\bm\tau}_{\cal U}$ the position of atom ${\cal U}$ 
within the unit cell.   
Further computing the Coulomb matrix in reciprocal space,
\begin{equation}
    V_{\mu\nu}(\bfq) = \sum_{\bfR} \ee^{\ii\bfq\cdot\bfR} V_{\mu\nu}(\bfR) = \sum_{\bfq} e^{i\bfq\cdot\bfR} \iint \frac{P_\mu(\bfr-{\bm\tau}_{\cal U})P_\nu(\bfrp-{\bm\tau}_{\cal V}-\bfR)}{|\bfr-\bfrp|} \textrm{d}\bfr \textrm{d}\bfrp \, ,
    \label{eq:Coulomb_matr_kspace}
\end{equation}
the RPA correlation energy can be evaluated using Eq.~\eqref{eq:cRPA},  where $\chi^0(i\omega)$ and $v$ should be interpreted as
their respective matrix forms represented in terms of the ABFs, as given by Eqs.~\eqref{eq:chi0_k_aux-expan} and \eqref{eq:Coulomb_matr_kspace}. 
Here, $\textrm{Tr}\left[AB\right] = \frac{1}{N_\bfk} \sum_{\mu,\nu,\bfq} \left[A_{\mu,\nu}(\bfq)B_{\nu,\mu}(\bfq)\right]$.
What is described above is a well-defined formalism
that yields reliable results, under the condition that the employed RI or LRI approximations are adequately accurate, and the singularity of 
the Coulomb matrix at $\bfq=0$ is properly treated. This above $\bfk$-space based formalism has been implemented in FHI-aims, 
and benchmark calculations have proven its numerical reliability. However,  the bottleneck step in RPA calculations, i.e.,
the evaluation of $\chi^0_{\mu\nu}(\bfq,\ii\omega)$, scales quartically with the number of
basis functions in the unit cell and quadratically with the number of $\bfk$ points, preventing its application to large systems.

To address this issue, here we reformulate the approach in real space, particularly 
taking advantage of the locality offered by NAO basis functions.
As usual, we start with the real-space imaginary-time expression of $\chi^0$, given by a simple product of the non-interacting Green's function
$G^0$,
\begin{equation}
    \chi^0(\textbf r,\textbf r',\ii\tau) = -\ii G^0(\textbf r,\textbf r',\ii\tau)G^0(\textbf r',\textbf r,-\ii\tau) \, .
    \label{eq:chi0_green}
\end{equation}
Within an AO basis framework, the KS wavefunctions in $\bfk$ space are given by
\begin{equation}
	\psi_{n\bfk}(\bfr) = \sum_{i} \sum_{\bfR} \ee^{\ii\bfk \cdot \bfR} c_{i,n}(\bfk) \varphi_i(\bfr-\bfR-{\bm \tau}_I) \, ,
\end{equation}
where $\varphi_i(\bfr)$ is a NAO sitting on atom $I$ (with ${\bm \tau}_I$ denoting its position within the unit cell), and $c_{i,n}(\bfk)$
are KS eigenvectors. 
The non-interacting Green's function $G^0(\ii\tau)$ in the imaginary-time domain can be expanded in terms of the NAOs as
\begin{equation}
	G^0(\bfr,\bfrp,\ii\tau) = \sum_{i,j}\sum_{\bfR_1,\bfR_2} \varphi_i(\bfr-\bfR_1-{\bm \tau}_I) G^0_{i,j}(\bfR_2-\bfR_1,\ii\tau) \varphi_{j}(\bfrp-\bfR_2-{\bm \tau}_J) \, .   
	\label{eq:green_func}
\end{equation}
\\
with
\begin{align}
	G^0_{i,j}(\textbf R,\ii\tau)= \left\{
	\begin{array}{ll}
	\displaystyle
	-\ii\frac{1}{N_\bfk}\sum_{n,\bfk}f_{n\textbf  k} c_{i,n}(\textbf  k) c_{j,n}^\ast (\textbf  k) \ee^{-\ii\textbf  k \cdot \textbf R} \ee^{-(\epsilon_{n,\textbf  k} - \mu) \tau} 
	 & \tau \le 0 \, , \\
	\displaystyle
	\ii\frac{1}{N_\bfk}\sum_{n,\textbf  k}(1-f_{n\textbf  k}) c_{i,n}(\textbf  k) c_{j,n}^\ast (\textbf  k) \ee^{-\ii\textbf  k \cdot \textbf R} \ee^{-(\epsilon_{n,\bfk} - \mu) \tau}  
	& \tau >0 \, .
	\end{array} \right.
	\label{eq:green_mat}
\end{align}
Here, $G^0_{i,j}(\bfR,\ii\tau)$ is the matrix form of $G^0(\ii\tau)$ represented in terms of NAOs, $\mu$ is the chemical potential,
and $\epsilon_{n,\bfk}$ and $f_{n\bfk}$ are KS orbital energies and occupation factors. Here, for simplicity, we assume $f_{n\bfk}$ equals 1
for occupied states and 0 for unoccupied ones. The situation of fractional occupations is more involved and will be discussed separately.

Plugging Eq.~\eqref{eq:green_func} into Eq.~\eqref{eq:chi0_green}, one has
\begin{align}
    \chi^0(\textbf r,\textbf r',\ii\tau)=-\ii   \sum_{i,j,k,l}\sum_{\bfR_1,\bfR_2, \bfR_3,\bfR_4} 
    & \varphi_i(\bfr-\bfR_1-{\bm \tau}_I) \varphi_k(\bfr-\bfR_3-{\bm \tau}_K) 
     G^0_{i,j}(\bfR_2-\bfR_1,\ii\tau)  \nonumber \\
    & G^0_{l,k}(\bfR_3-\bfR_4,-\ii\tau) \varphi_{j}(\bfrp-\bfR_2-{\bm \tau}_J)\varphi_{l}(\bfrp-\bfR_4-{\bm \tau}_L)
    \label{eq:chi0_NAOs}
\end{align}
where ${\bm \tau}_K$ and ${\bm \tau}_L$ denote the positions of the atom $K$ and $L$, on which the basis function $\varphi_k(\bfr)$
and $\varphi_l(\bfr)$ are sitting, respectively.
The key idea here is to derive a more compact representation of $\chi^0(\textbf r,\textbf r',\ii\tau)$ in terms of the ABFs, i.e.,
\begin{equation}
\chi^0(\bfr,\bfrp,\ii\tau)=\sum_{\mu \in \mathcal{U},\nu \in \mathcal{V}} \sum_{\textbf{R}_1,\textbf{R}_2}P_\mu(\bfr-\textbf{R}_1-{\bm \tau}_{\mathcal{U}})\chi^0_{\mu,\nu}(\textbf{R}_2-\textbf{R}_1,\ii\tau)P_{\nu}(\bfrp-\textbf{R}_2-{\bm \tau}_\mathcal{V})
\label{eq:chi0_ABFs}
\end{equation}
with ${\cal U}$ and ${\cal V}$ denoting the atoms on which the ABFs $P_\mu(\bfr)$ and $P_\nu(\bfr)$ are sitting, and ${\bm \tau}_{\cal U}$ and ${\bm \tau}_{\cal V}$ their respective atomic positions within the unit cell. 
To this end, we apply the LRI approximation here,  which in essence expands the product of two NAOs in terms of the ABFs sitting
on the two atoms on which the two NAOs are centering, i.e.,
\begin{align}
		 &  \varphi_{i}(\bfr-\bfR_1-{\bm \tau}_I)\varphi_{k}(\bfr-\bfR_3-{\bm \tau}_K)     \nonumber \\
         \approx  &  \sum_{\mu \in I} C_{i(\bfR_1),k(\bfR_3)}^{\mu(\bfR_1)} P_\mu(\bfr-\bfR_1-{\bm \tau}_I)  + 
		   \sum_{\mu \in K} C_{i(\bfR_1),k(\bfR_3)}^{\mu(\bfR_3)}P_\mu(\bfr-\bfR_3-{\bm \tau}_K) \nonumber \\
	   = & \sum_{\mu \in I} C_{i(\bfo),k(\bfR_3-\bfR_1)}^{\mu(\bfo)} P_\mu(\bfr-\bfR_1-{\bm \tau}_I) + 
               \sum_{\mu \in K} C_{i(\bfR_1-\bfR_3),k(\bfo)}^{\mu(\bfo)} P_\mu(\bfr-\bfR_3-{\bm \tau}_K) \, .
		    \label{eq:LRI}
\end{align}
Here we follow the notation adopted in Ref.~\citenum{Ren/etal:2021}, 
whereby $\tilde{C}_{i(\bfR_1),k(\bfR_3)}^{\mu(\bfR_1)}$ denote the two-center expansion coefficients 
with the lattice vector in parentheses indicating the unit cell to which the basis function
belongs. Furthermore,  $\mu \in I$ ($\mu \in K$) in Eq.~\eqref{eq:LRI} signifies that the summation over the ABFs is restricted to those centering at the atom $I$ ($K$). 
The second equation of Eq.~\eqref{eq:LRI} follows from the translational symmetry of 
the periodic system, which requires that 
      $C_{i(\bfR_1),k(\bfR_3)}^{\mu(\bfR_1)}=C_{i(\bfzero),k({\bfR}_3-{\bfR}_1)}^{\mu(\bfzero)}$, with $\bfzero$ here denoting
the unit cell at the origin. This implies that the expansion coefficients only depend on one independent lattice vector.

Now, by equalizing Eq.~\eqref{eq:chi0_NAOs} with Eq.~\eqref{eq:chi0_ABFs}, and utilizing Eq.~\eqref{eq:LRI}, it is somewhat lengthy but otherwise
straightforward to show that the matrix form of $\chi^0(i\tau)$ in real space is given as follows,
\begin{align}
	\label{eq:chi0_matrix}
	\chi_{\mu,\nu}^0(\textbf{R},\ii\tau) 
	&=-\ii \Bigg[\sum_{i \in \mathcal{U}}\sum_{k \in K,\textbf{R}_1}C_{i(\textbf{0}),k(\textbf{R}_1)}^{\mu (\textbf{0})}\left(  M^{\nu}_{i,k}(\textbf{R}_1,\textbf{R},\ii\tau)+ M^{\nu*}_{i,k}(\textbf{R}_1,\textbf{R},-\ii\tau)\right.\nonumber\\
	&\qquad\qquad\qquad\qquad\qquad\left.+ Z^{\nu}_{i,k}(\textbf{R}_1,\textbf{R},\ii\tau)+ Z^{\nu*}_{i,k}(\textbf{R}_1,\textbf{R},-\ii\tau) \right)\Bigg] \nonumber\\
 	&= -\ii\sum_{i \in \mathcal{U}}\sum_{k \in K,\textbf{R}_1}C_{i(\textbf{0}),k(\textbf{R}_1)}^{\mu (\textbf{0})}O^{\nu}_{i,k}(\textbf{R}_1,\textbf{R},\ii\tau),
\end{align}
where
\begin{align}
	\label{eq:chi0_parts_sum}
   &O^{\nu}_{i,k}(\textbf{R}_1,\textbf{R},i\tau)=M^{\nu}_{i,k}(\textbf{R}_1,\textbf{R},\ii\tau)+ M^{\nu*}_{i,k}(\textbf{R}_1,\textbf{R},-\ii\tau)\nonumber \\
	 &\qquad \qquad\qquad\quad+ Z^{\nu}_{i,k}(\textbf{R}_1,\textbf{R},\ii\tau)+ Z^{\nu*}_{i,k}(\textbf{R}_1,\textbf{R},-\ii\tau)
  \end{align}
and
\begin{align}
     \label{eq:chi0_parts}
	&M^{\nu}_{i,k}(\textbf{R}_1,\textbf{R},\ii\tau)=\sum_{j \in \mathcal{V}}G_{i,j}(\textbf{R},\ii\tau)N^{\nu}_{j,k}(\textbf{R}_1,\textbf{R},\ii\tau)\nonumber \\
	&Z^{\nu}_{i,k}(\textbf{R}_1,\textbf{R},\ii\tau)=\sum_{j \in \mathcal{V}}G_{j,k}(\textbf{R}_1-\textbf{R},-\ii\tau)X^\nu_{i,j}(\textbf{R},\ii\tau)
 \end{align}
 with the intermediate quantities $N^{\nu}_{j,k}(\textbf{R}_1,\textbf{R},\ii\tau)$ and $X^\nu_{i,j}(\textbf{R},\ii\tau)$ defined as
 \begin{align}
 \label{eq:chi0_NX_intermidate}
 	&N^{\nu}_{j,k}(\textbf{R}_1,\textbf{R},\ii\tau)=\sum_{l\in L,\textbf{R}_2}C_{j(\textbf{0}),l(\textbf{R}_2-\textbf{R})}^{\nu (\textbf{0})}G_{l,k}(\textbf{R}_1-\textbf{R}_2,-\ii\tau)\nonumber \\
	&X^\nu_{i,j}(\textbf{R},\ii\tau)=\sum_{l\in L,\textbf{R}_2}C_{j(\textbf{0}),l(\textbf{R}_2-\textbf{R})}^{\nu (\textbf{0})}G_{i,l}(\textbf{R}_2,\ii\tau) \, .
\end{align}
In deriving the above equations, symmetry properties of the expansion coefficients and index swapping have been used. Details of the derivations 
are presented in the Appendix~\ref{app:equation_derivation}. Eqs. (\ref{eq:chi0_matrix}-\ref{eq:chi0_NX_intermidate}) are the key underlying equations on which the
low-scaling algorithm is based, which will be discussed in the next section.


So far, we have constructed the response function matrix in terms of ABFs in the real-space imaginary-time domain. 
To compute the RPA correlation energy [Eq.~\eqref{eq:cRPA}],
it is more convenient to work in the $\bfk$ space and imaginary frequency domain.
To this end, Fourier transforms from the real to reciprocal spaces, and from the imaginary time to imaginary frequency domains are 
sequentially performed for the response function matrix.
Considering the symmetry property of $\chi^0$ in time and frequency, i.e., 
$\chi^0(\textbf r, \textbf r' , \ii\omega)=\chi(\textbf r', \textbf r , -\ii\omega)$,
$\chi^0(\textbf r, \textbf r' , \ii\tau)=\chi^0(\textbf r', \textbf r , -\ii\tau)$, and $\chi^0(\bfR,\ii\tau) =\chi^0(-\bfR,\ii\tau)$, 
the time-to-frequency Fourier transform between the complex axes is simplified to a cosine transformation 
including an additional factor of $-\ii$,\cite{Kaltak/Merzuk/Kresse:2014-low-scalng,Kaltak/etal:2014b,Rieger/etal:1999}
\begin{equation}
    \label{eq:cos_f2t}
    \chi^0_{\mu,\nu}(\bfR, \ii\omega_k)=-\ii\sum_{j=1}^{N} \gamma_{jk}\chi^0_{\mu,\nu}(\bfR , \ii\tau_j)\, cos(\tau_j \omega_k)\, . 
\end{equation}
Here we adopt the nonuniform imaginary-time $\{\ii\tau_j\}_{j=1}^{N_{\tau}}$ and frequency $\{\ii\omega_k\}_{k=1}^{N_{\omega}}$ minimax grids from CP2K\cite{CP2Kweb}, which have been proven to be accurate\cite{wilhelm/dorothea:2021}. The coefficients $\gamma_{jk}$ are determined using $L^2$ minimization \cite{Kaltak/Merzuk/Kresse:2014-low-scalng} during program run.
Once the real-space imaginary-frequency $\chi^0$ matrix is obtained from Eq.~\eqref{eq:cos_f2t}, it is further transformed to the reciprocal space
straightforwardly,
\begin{equation}
    \label{eq:FT_r2k}
    \chi^0_{\mu,\nu}(\textbf q, \ii\omega)=\sum_{\textbf R}\ee^{\ii\bfq \cdot\bfR}\chi^0_{\mu,\nu}(\bfR,\ii\omega) \, .
\end{equation}

To facilitate the computation of RPA correlation energy, we further introduce an intermediate quantity, i.e., the product
of $\chi^0$ and $V$ matrices,
\begin{equation}
	\Pi(\bfk, \ii\omega)=\chi^0(\bfk, \ii\omega)V(\bfk)\, .
	\label{eq:pi_matrix}
\end{equation}
The RPA correlation energy for periodic systems per unit cell can be finally obtained as
\begin{align}
	E^\textrm{RPA}_\textrm{c}&=\frac
    {1}{2\pi}\frac{1}{N_\bfk}\sum_\bfq \int^{\infty}_0 \textrm{d}\omega\, \textrm{Tr}\left[\textrm{ln}(1-\Pi( \bfq, \ii\omega))+\Pi(\bfq, \ii\omega)\right] \nonumber \\
	&=\frac
    {1}{2\pi} \frac{1}{N_\bfk}\sum_\bfq \int^{\infty}_0 \textrm{d}\omega\, \textrm{ln}\left[\textrm{det}(1-\Pi(\bfq, \ii\omega))\right]+\textrm{Tr}[\Pi(\bfq, \ii\omega)],
	\label{eq:cRPA_use_pi}
\end{align}
where the property $\textrm{Tr}[\textrm{ln}(A)]=\textrm{ln}[\textrm{det}(A)]$ is used.

\section{ Implementation Details}
\label{sec:implementation}
\begin{figure}[htbp] 
\includegraphics[width=0.8\textwidth]{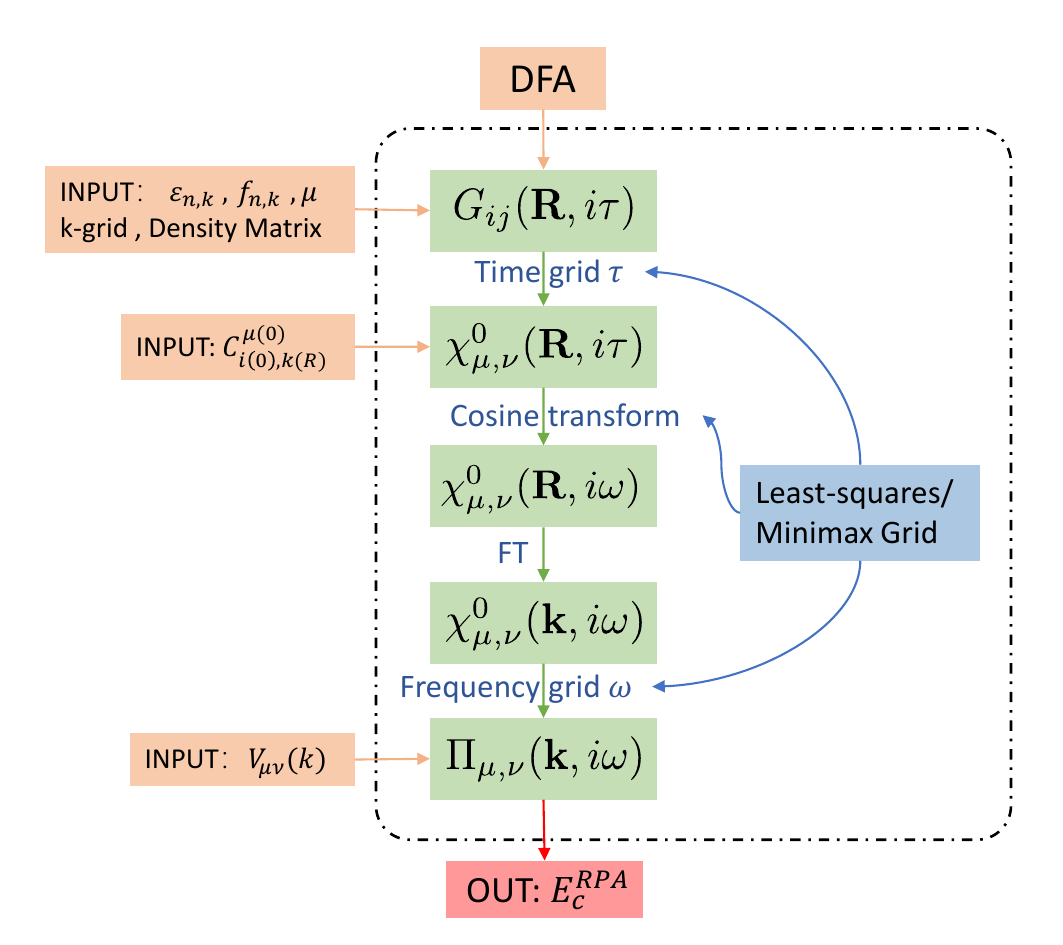}
\centering
\caption{Major steps in the computation of the RPA correlation energy in the present algorithm: i) Calculate the non-interacting Green's function
within the NAO basis set based on a preceding DFA calculation; ii) contract two Green's functions to construct $\chi^0_{\mu,\nu}(\bfR, \ii\tau)$ using LRI; iii) perform the cosine transformation to obtain $\chi^0_{\mu,\nu}(\bfR, \ii\omega)$ according to Eq.~\eqref{eq:cos_f2t}; iv) Fourier transform (FT) $\chi^0_{\mu,\nu}(\bfR, i\omega)$ to $\chi^0_{\mu,\nu}(\bfk, \ii\omega$); v) calculate $\Pi(\bfk,\ii\omega)$ according to Eq.~\eqref{eq:pi_matrix}; vi) calculate $E_c^\text{RPA}$ via an integration
over imaginary frequencies and summation over BZ [Eq.~\eqref{eq:cRPA_use_pi}].} 
\label{fig:flowchart}
\end{figure}

The major steps of computing $E_c^\text{RPA}$ within the NAO basis sets are illustrated in Fig.~\ref{fig:flowchart}. 
After a preceding self-consistent
KS calculation with a lower-rung density functional approximation (DFA), one obtains the KS orbitals and orbital energies. The ABFs have been generated beforehand, and so do the LRI coefficients $C_{i(\bfo),j(\bfR)}^{\mu(\bfo)}$ (Eq.~\eqref{eq:LRI}) 
and the Coulomb matrix $V_{\mu\nu}(\bfk)$ (Eq.~\eqref{eq:Coulomb_matr_kspace}). 
These quantities have been available previously and used
in periodic hybrid functional \cite{Lin/Ren/He:2020,Lin/Ren/He:2021} and $G_0W_0$ calculations \cite{Ren/etal:2021}.
The implementation in the present work begins with the evaluation of real-space imaginary-time independent-particle Green's function 
$G^0(\bfR,\ii\tau)$ using 
eigenvalues and eigenvectors generated using a NAO-based DFT code [cf. Eq.~\eqref{eq:green_mat}].
After calculating and storing the $G^0_{ij}(\bfR,\ii\tau)$, the real-space imaginary-time response function matrix 
$\chi^0_{\mu\nu}(\bfR,\ii\tau)$ can be evaluated, which is usually the rate-determining step throughout the whole RPA computation.  The inputs needed for this step are the Green's function matrix $G^0_{ij}(\bfR,\ii\tau)$ and
the LRI coefficients $C_{i(\bfo),j(\bfR)}^{\mu(\bfo)}$. After $\chi^0_{\mu\nu}(\bfR,\ii\tau)$ is obtained, it is relatively straightforward to
convert it to $\chi^0_{\mu\nu}(\bfk,\ii\omega)$ via the cosine transform and Fourier transform successively. The cosine transform benefits
from the recently developed efficient minimax quadrature grids \cite{Kaltak/Klimes/Kresse:2014,PeitaoLiu/etal:2016,Wilhelm/etal:2016}. With $\chi^0_{\mu\nu}(\bfk,\ii\omega)$, one can multiply it with the Coulomb matrix $V_{\mu,\nu}(\bfk)$ to obtain $\Pi_{\mu,\nu}(\bfk,\ii\omega)$, 
and finally compute the RPA correlation energy via Eq.~\eqref{eq:cRPA_use_pi}.

\begin{algorithm}[H]
    \caption{\label{alg:algorithm_N2} Loop structure of evaluating $\chi^0_{\mu\nu}(\bfR,\ii\tau)$.  $<\mathcal U(\textbf 0),\mathcal V(\textbf R)>$ denotes
    an atomic pair with atom ${\mathcal U}$ in the unit cell at origin and atom ${\mathcal V}$ in the unit cell $\bfR$. The symbol $\mathcal N[\mathcal U]$ represents the set of neighboring atoms of atom $\mathcal U$, and $K(\textbf R_1) \in \mathcal{N}[\mathcal U(\textbf 0)]$ means
    that the atom $K$ in unit cell $\bfR_1$ is in the neighborhood of the atom $\mathcal U$ in the unit cell at the origin.}
    \begin{algorithmic}[1]
        \ForAll { $\tau $}
            \ForAll { $\textbf R$}
                \ForAll { $<\mathcal U(\textbf 0),\mathcal V(\textbf R)>$}
                    \ForAll{$L(\textbf R_2) \in \mathcal{N}[\mathcal V(\textbf R)]$} \label{alg1:loop_atp_begin}
                        \State  Calculate  $X^\nu_{i,j}(\textbf{R}_1,\textbf{R},\ii\tau)$ \hfill [cf. Eq.~\eqref{eq:chi0_NX_intermidate}]
                    \EndFor
                    \ForAll{$K(\textbf R_1) \in \mathcal{N}[\mathcal U(\textbf 0)]$}
                        \ForAll{$L(\textbf R_2) \in \mathcal{N}[\mathcal V(\textbf R)]$}
                            \State Calculate  $N^{\nu}_{j,k}(\textbf{R}_1,\textbf{R},\ii\tau)$ \hfill [cf. Eq.~\eqref{eq:chi0_NX_intermidate}]
                        \EndFor
                        \State Calculate  $M^{\nu}_{i,k}(\textbf{R}_1,\textbf{R},\ii\tau)$, $Z^{\nu}_{i,k}(\textbf{R}_1,\textbf{R},\ii\tau)$ \hfill [cf. Eq.~\eqref{eq:chi0_parts}]
                        \State Calculate $O^{\nu}_{i,k}(\textbf{R}_1,\textbf{R},\ii\tau)$ \hfill [cf. Eq.~\eqref{eq:chi0_parts_sum}]
						\State Calculate $\chi^0_{\mu,\nu}(\textbf{R},\ii\tau)\mathrel{{+}{=}}C_{i(\textbf{0}),k(\textbf{R}_1)}^{\mu (\textbf{0})}O^{\nu}_{i,k}(\textbf{R}_1,\textbf{R},\ii\tau)$ \hfill [cf. Eq.~\eqref{eq:chi0_matrix}]
                    \EndFor
					\label{alg1:loop_atp_end}
                \EndFor
            \EndFor
        \EndFor
    \end{algorithmic}
    \label{fig:algorithm}
\end{algorithm}

The essential point of the present work is to reduce the computational scaling of evaluating $\chi^0_{\mu,\nu}(\bfR, \ii\tau)$. Algorithm~\ref{fig:algorithm} illustrates the loop structure of computing $\chi^0_{\mu,\nu}(\bfR, \ii\tau)$ based on Eqs.~(\ref{eq:chi0_matrix}-\ref{eq:chi0_NX_intermidate}). The outermost loop goes over all time grid points $\{\tau_j\}_{j=1}^{N_{\tau}}$, under which one further goes through all lattice vectors 
$\{ \bfR\}$ within the BvK supercell. For each $(\tau, \bfR)$ pair, the 
whole $\chi^0_{\mu,\nu}(\bfR, \ii\tau)$ matrix is decomposed into blocks associated with 
individual atomic pairs $<\mathcal U(\textbf 0),\mathcal V(\textbf R)>$ on which
the ABFs $P_\mu$ and $P_\nu$ are located, respectively. Computing these blocks $\chi^0_{\mu\in \mathcal U, \nu\in \mathcal V}$ separately for each
atomic pair  $<\mathcal U(\textbf 0),\mathcal V(\textbf R)>$ and assembling them up, one obtains the entire $\chi^0_{\mu,\nu}$ matrix. 
Obviously, for a given lattice vector $\bfR$, the number of such atomic pairs scales as $N_{\rm at}^2$ where $N_{\rm at}$ is the number of atoms in a unit cell. 

Now, inside the loop over the atomic pair $<\mathcal U(\textbf 0),\mathcal V(\textbf R)>$, one still needs to go through atom $K$ in the
unit cell of $\bfR_1$, and atom $L$ in the unit cell of $\bfR_2$, in order to compute intermediate quantities such as $N^{\nu}_{j,k}(\textbf{R}_1,\textbf{R},\ii\tau)$ and $X^\nu_{i,j}(\textbf{R},\ii\tau)$ as defined in Eq.~\eqref{eq:chi0_NX_intermidate}, and
finally $\chi^0_{\mu\in \mathcal U, \nu\in \mathcal V}$. The key point here is that the atom $K(\bfR_1)$ has to be the neighboring
atom of the atom $\mathcal U(\textbf 0)$, and $L(\bfR_2)$ has to be the neighboring atom of  $\mathcal V(\textbf R)$. Outside the
neighborhood region, the LRI expansion coefficients will be zero (or insignificantly small) and the $K$, $L$ atoms there  
will not contribute. For a finite periodic system, the number of
neighboring atoms of a given reference atom is determined by the spatial range (cutoff radii) of NAO basis functions, 
and does not keep increasing with 
size and complexity of the unit cell. This means that, in our algorithm, the computational cost required for a block
of response function matrix associated with an atomic pair $<\mathcal U(\textbf 0),\mathcal V(\textbf R)>$ approaches a constant as the size
of the system (unit cell) grows. Thus the entire computational cost scales as $N_{\rm at}^2 N_\bfR N_\tau$ or $N_{\rm at}^2 N_\bfk N_\tau$ where $N_\bfR$ is
the number of unit cells in the BvK supercell, usually set equal to $N_\bfk$, and $N_\tau$ is the
number of imaginary time grid points. Note that in practical converged calculations, the size of unit cells (i.e., $N_{\rm at}$) is not independent
of $N_\bfR$ or $N_\bfk$; large unit cells usually mean that one can use fewer $\bfk$ points or even a single $\Gamma$ point in the
calculations. Thus, $N_{\rm at}\times N_\bfk$ can be roughly considered as a constant for a given type of system, and furthermore $N_\tau$ does not
increase noticeably with system size. Thus, the above described algorithm is \textit{de facto} quadratic scaling for 
evaluating the response function matrix.

\begin{algorithm}[H]
    \caption{\label{alg:algorithm2} Refined algorithm for evaluating $\chi^0_{\mu\nu}(\bfR,\ii\tau)$ whereby the Green's-function-based screening is incorporated. This will reduce the number of loops in Algorithm~\ref{alg:algorithm_N2} for$<\mathcal U(\textbf 0),\mathcal V(\textbf R)>$, 
    as well as the matrix multiplications inside the loop, to varying degrees.}
    \begin{algorithmic}[1]
		\ForAll { $\tau $}
            \ForAll { $\textbf R$}
                \ForAll { $<\mathcal U(\textbf 0),\mathcal V(\textbf R)>$}
                    \ForAll{$L(\textbf R_2) \in \mathcal{N}[\mathcal V(\textbf R)]$} 
						\If {${ \max \{ |G_{i\in \mathcal U(\textbf 0), l\in L(\textbf R_2)} (\bfR_2, \ii\tau)| \} >\eta_G}$}
                        	\State  Calculate  $X^\nu_{i,j}(\textbf{R}_1,\textbf{R},\ii\tau)$ \hfill [cf. Eq.~\eqref{eq:chi0_NX_intermidate}]
						\EndIf
					\EndFor
                    \ForAll{$K(\textbf R_1) \in \mathcal{N}[\mathcal U(\textbf 0)]$}
						\If {${ \max \{ |G_{i\in \mathcal U(\textbf 0), j\in \mathcal V(\textbf R)} (\bfR, \ii\tau)| \} >\eta_G}$}
                        \ForAll{$L(\textbf R_2) \in \mathcal{N}[\mathcal V(\textbf R)]$}
							\If {${ \max \{ |G_{l\in L(\textbf R_2), k\in K(\textbf R_1)} (\bfR_1-\bfR_2, -\ii\tau)| \} >\eta_G}$}
								\State Calculate  $N^{\nu}_{j,k}(\textbf{R}_1,\textbf{R},\ii\tau)$ \hfill [cf. Eq.~\eqref{eq:chi0_NX_intermidate}]
							\EndIf
						\EndFor
						\State Calculate  $M^{\nu}_{i,k}(\textbf{R}_1,\textbf{R},\ii\tau)$\hfill [cf. Eq.~\eqref{eq:chi0_parts}]
						\EndIf

						\If {${ \max \{ |G_{j\in \mathcal V(\textbf R), k\in K(\textbf R_1)} (\bfR_1-\bfR, -\ii\tau)| \} >\eta_G}$}
                        	\State Calculate   $Z^{\nu}_{i,k}(\textbf{R}_1,\textbf{R} ,\ii\tau)$ \hfill [cf. Eq.~\eqref{eq:chi0_parts}]
						\EndIf 
						\State Calculate $O^{\nu}_{i,k}(\textbf{R}_1,\textbf{R},\ii\tau)$ \hfill [cf. Eq.~\eqref{eq:chi0_parts_sum}]
						\State Calculate $\chi^0_{\mu,\nu}(\textbf{R},\ii\tau) \mathrel{{+}{=}} C_{i(\textbf{0}),k(\textbf{R}_1)}^{\mu (\textbf{0})}O^{\nu}_{i,k}(\textbf{R}_1,\textbf{R},\ii\tau)$ \hfill [cf. Eq.~\eqref{eq:chi0_matrix}]
                    \EndFor
                \EndFor
            \EndFor
        \EndFor
    \end{algorithmic}
    \label{fig:algorithm_green_threshold}
\end{algorithm}
In the above analysis of the scaling behavior of Algorithm~\ref{alg:algorithm_N2}, the sparsity of the Green's function matrix $G^0_{ij}(\bfR,\ii\tau)$ itself was not taken into account.
In fact, $G^0_{ij}(\bfR,\ii\tau)$ at $\tau \rightarrow 0^{-}$ corresponds to the reduced one-electron density matrix, 
which is known to decay exponentially for
insulating systems, and polynomially for metallic systems, as the distance $|\bfR+{\bm \tau}_j -{\bm \tau}_i|$ between the centers of 
atomic orbitals $i$ and $j$ gets large. 
Exploiting this property, one can envision that the number of  relevant atomic pairs $<\mathcal U(\textbf 0),\mathcal V(\textbf R)>$ 
does not grow quadratically with respect to the unit cell size any longer, 
but rather linearly. This suggests that a refined algorithm that accounts for
the sparsity of the Green's function $G^0_{ij}(\bfR,\ii\tau)$ will become asymptotically linear-scaling. In practical implementation of this concept,
one can introduce a screening threshold $\eta_G$, whereby, if the maximal element of the Green's function matrix associated with an atomic pair 
$<{\mathcal U}(\textbf 0),{\mathcal V}(\textbf R)>$ is smaller than $\eta_G$, i.e.,
\begin{equation}
    \max \{ |G_{i\in \mathcal U(\textbf 0), j\in \mathcal V(\textbf R)} (\bfR, \ii\tau)| \} <\eta_G \, ,
\end{equation}
then this atomic pair will be discarded in the evaluation of the $\chi^0$ matrix. Algorithm~\ref{alg:algorithm2} illustrates the basic idea behind this refined 
scheme, leading to an asymptotically linear-scaling algorithm for evaluating $\chi^0(\bfR,\ii\tau)$. 
In Sec.~\ref{sec:gf_screening} we will demonstrate the effect of filtering out the
atomic pairs with zero or sufficiently small Green's function matrix elements.

The above-described algorithm for low-scaling RPA correlation energy calculations has been implemented in a standalone 
library called LibRPA, which is currently accessible from GitHub \cite{LibRPAweb}. So far, LibRPA has been
interfaced with two NAO-based first-principles code packages -- the all-electron FHI-aims code \cite{Blum/etal:2009} and the
pseudopotential-based ABACUS code \cite{Li/Liu/etal:2016}. Interfacing with other DFT codes that employ NAOs should be
straightforward,
if the necessary inputs as shown in Fig.~\ref{fig:flowchart} can be provided.



\section{Results}
\label{sec:results}

In this section, we set out to benchmark the performance of the low-scaling RPA algorithm and implementation as described in previous sections, 
for selected insulating and semiconducting systems. Both the numerical accuracy and the efficiency of the
implementation will be examined here. Regarding the efficiency, we will particularly check the practical scaling behavior of 
the computational cost with respect to the system size.

\subsection{Accuracy of RPA correlation energy}
\label{sec:accuracy}
We first examine the numerical accuracy of our low-scaling algorithm. To this end, we compare the RPA correlation energies as calculated by LibRPA
with those produced by the conventional $\bfk$-space implementation in FHI-aims. The conventional implementation is also based on
LRI, but the key operations are performed in $\bfk$-space, without exploiting the sparsity of the LRI coefficients and the Green's function.
This leads to a $\mathcal{O}(N^4)$ scaling for calculating the response function matrix $\chi^0(\bfk, i\omega)$. The algorithm and
implementation details follow closely the periodic $G_0W_0$ implementation as described in Ref.~\cite{Ren/etal:2021}. Production calculations
based on such a conventional implementation have been reported in Refs.~\cite{IgorZhang/etal:2019,Yang/Ren:2022}.



\begin{table}[htbp]
\caption{RPA correlation energies for several semiconductors as calculated by the real-space low-scaling algorithm as implemented in LibRPA and 
by the $\bfk$-space algorithm as implemented in FHI-aims. The FHI-aims ``\textit{tight}'' NAO basis sets are used for all semiconductors. For 
 for some of the systems (Si, BN, and MgO), the results obtained using loc-NAO-VCC-3Z basis sets are also presented. 
 A $4\times 4 \times 4$ $\bfk$ grid is adopted
for all calculations. For FHI-aims calculations, a modified Gauss-Legendre frequency quadrature grid with 80 points is used,
and for LibRPA calculations, minimax grids with 18 points for both time and frequency are used. Frozen-core approximation is used
for all calculations.}
\centering
\begin{tabular}{c c c c c c}
\toprule
&         & Basis set        & FHI-aims (eV)     & LibRPA (eV)    &Difference (meV)\\
\hline
&\multirow{2}{*}{Si} & \textit{tight}   &-15.836307	&-15.836364	&0.0574 \\
&         & loc-NAO-VCC-3Z	 &-18.321370	&-18.321399	&0.0291 \\
&\multirow{2}{*}{BN}&\textit{tight}   &-27.816262	&-27.816345	&0.0831 \\
&	      & loc-NAO-VCC-3Z    &-29.428039	&-29.428047	&0.0085 \\
&\multirow{2}{*}{MgO }&\textit{tight} &-11.55797	    &-11.557899	&-0.0704 \\
&         & loc-NAO-VCC-3Z    &-9.991418	    &-9.991505	&0.0877 \\
&SiC      & \textit{tight}   			 &-64.421342	&-64.421246	&-0.0955 \\
&GaAs     & \textit{tight}   			 &-17.562131	&-17.562132	&0.0007 \\
&ZnO      & \textit{tight}   			 &-50.186035	&-50.18603	    &-0.0059 \\

\hline
\label{table:cRPA_accuracy}
\end{tabular}
\end{table}

Table~\ref{table:cRPA_accuracy} presents the RPA correlation energies of several semiconductors, as obtained using the real-space, imaginary-time
algorithm as implemented in LibRPA, in comparison with those obtained using the conventional $\bfk$-space algorithm as implemented in FHI-aims. The same computational settings (basis sets, $\bfk$ grid, RI and frozen-core approximations) are used in both FHI-aims and 
LibRPA calculations. Sufficiently many imaginary frequency points (and imaginary time points in case of LibRPA)  are used 
in both types of calculations. 
Table~\ref{table:cRPA_accuracy} indicates that the LibRPA implementation produces nearly identical results as the conventional $\bfk$-space
implementation in FHI-aims. The difference in total RPA correlation calculations for all tested systems are below 0.1 meV. This holds for
both the FHI-aims-2009 \cite{Blum/etal:2009} (``\textit{tight}'' setting) and the localized variant of the NAO-VCC-$n$Z  \cite{IgorZhang/etal:2013} (denoted as loc-NAO-VCC-$n$Z) basis sets.
This is a remarkably high numerical precision, which validates the correctness of the proposed low-scaling RPA algorithm and the actual implementation carried out in LibRPA. 


\subsection{The scaling behavior of the real-space algorithm}
\label{sec:scaling_test}
\begin{figure}[htbp]
    \centering
    \subfigure[]
    {
        \label{fig:scaling-atom-a}
        \includegraphics[width=0.47\textwidth]{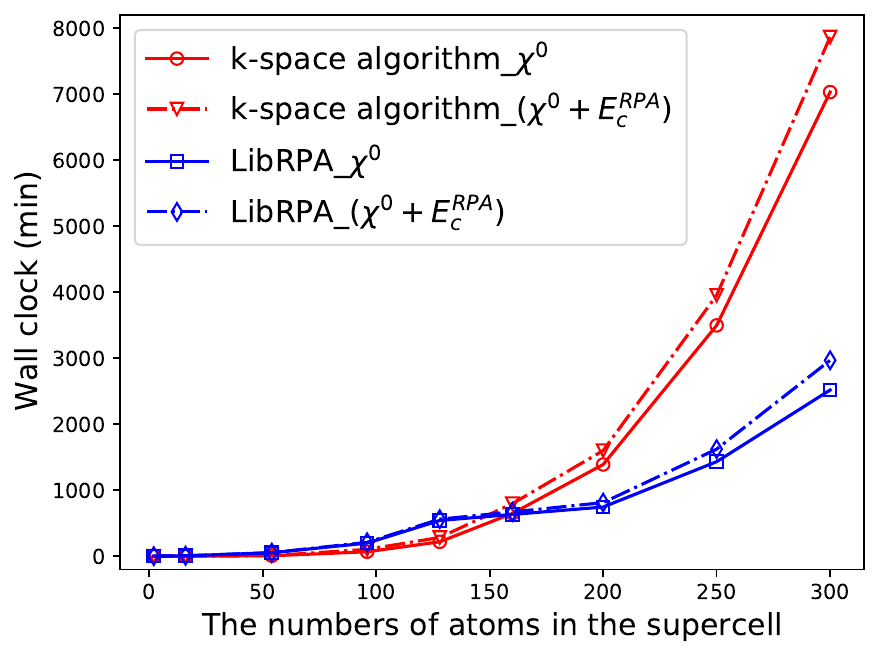 }
    }
    \subfigure[]
    {
        \label{fig:scaling-atom-b}
        \includegraphics[width=0.47\textwidth]{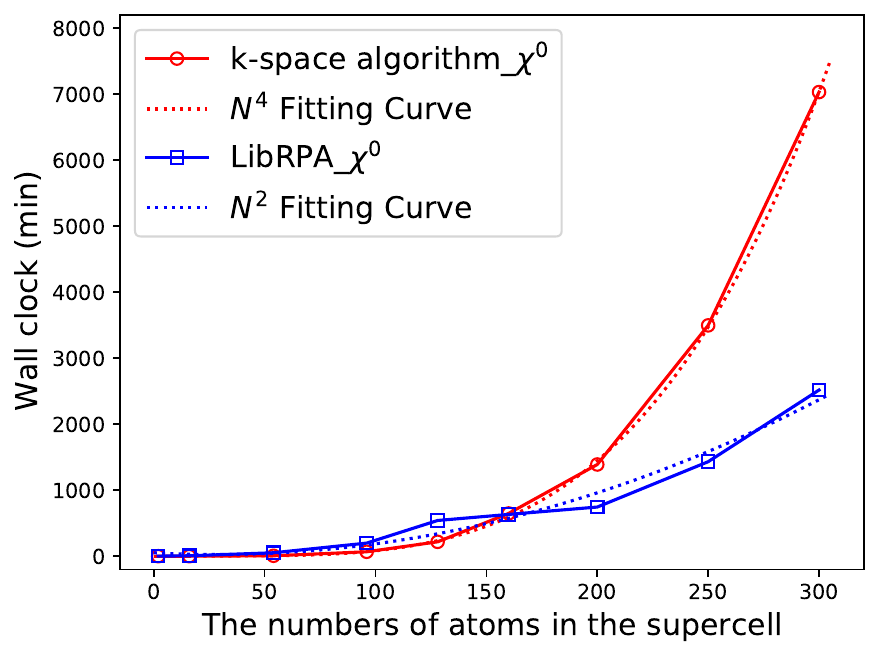 }
    }
    \caption{Scaling behavior of the computation time as a function of system size (number of atoms in the supercell) 
       for both the real-space low-scaling algorithm as implemented in LibRPA and the conventional $\bfk$-space algorithm
       as implemented in FHI-aims.  The test system is C diamond with increasing unit cell sizes. The loc-NAO-VCC-3Z basis set
       and a single ($\Gamma$-only) $\bfk$ point is used in the calculations.  For FHI-aims calculations, a modified Gauss-Legendre frequency quadrature grid with 40 points are used, whereas for LibRPA calculations, minimax grids with 12 points for both time and frequency are used. The vertical axis represents the time measured or converted to the usage of a compute node with 64 CPU cores (for large systems more than one
       compute node is needed to run the calculations, and in these cases the reported timing is rescaled as if the calculations were done on one node). Left panel: the timings for both evaluating $\chi^0$ matrix (solid lines) and the total RPA calculation $\chi^0+E_\text{c}^\text{RPA}$ 
       (dash-dotted curves) are presented. Right panel: $\mathcal{O}(N^4)$- and $\mathcal{O}(N^2)$-scaling curves (dotted lines) are added by fitting to the data 
       of the  conventional and low-scaling algorithms for evaluating the $\chi^0$ matrix, respectively.} 
           \label{fig:scaling-atom}
\end{figure}

With the validity of the algorithm and the correctness of the implementation being established, we now check the actual scaling behavior of 
our implementation with respect to system size. Specifically, we carried out RPA calculations for carbon diamond crystals with increasing supercell
size. In Fig.~\ref{fig:scaling-atom}, the computational timings of the low-scaling algorithm as implemented in LibRPA (blue curves) and the conventional $\bfk$-space algorithm (red curves) as implemented in FHI-aims are presented as a function of the supercell size (number of atoms). 
In the left panel of Fig.~\ref{fig:scaling-atom}, both the timings for constructing the response function matrix $\chi^0$ (solid lines), 
and the total computation times including, in addition to the construction of $\chi^0$ matrix, the rest of calculations all the way 
up to the final evaluation of $E_\text{c}^\text{RPA}$ (dash-dotted lines), 
are presented. The settings of the
computational parameters are chosen such that the two series of calculations yield nearly identical RPA correlation energies for the same system.
In the right panel of Fig.~\ref{fig:scaling-atom}, we only presented the timings for evaluating the $\chi^0$ matrix, but added 
the $\mathcal{O}(N^4)$ and $\mathcal{O}(N^2)$ fitting curves (dotted) for the computational times.

As expected, the computational cost for the $\bfk$-space implementation for evaluating the $\chi^0$ matrix 
shows a roughly $\mathcal{O}(N^4)$ scaling behavior with respect to system size
$N$ (here $N$ being the number of atoms in the supercell). In contrast, the real-space implementation in LibRPA shows a significantly
reduced scaling behavior, but with a larger prefactor. The crossing point occurs at system size of about 160 C atoms, and after that the
low-scaling algorithm starts to gain supremacy. In the benchmark tests presented in Fig.~\ref{fig:scaling-atom}, the Green's-function-based screening was 
not turned on, and thus the computational cost should ideally follow a $\mathcal{O}(N^2)$ scaling behavior as described in Algorithm~\ref{alg:algorithm_N2}. 
However, due to the fact that we have to increase the compute nodes for the larger systems and the complication arising from parallel efficiency,
some of the data points deviate from the ideal $\mathcal{O}(N^2)$ behavior. However, an overall $\mathcal{O}(N^2)$ scaling behavior is observable.

Furthermore, from Fig.~\ref{fig:scaling-atom}, one can see that for system size below 300 atoms, the computational cost of evaluating the response 
function matrix dominates. The rest of the calculations for evaluating the RPA correlation energy, though involving $\mathcal{O}(N^3)$ matrix multiplication and
Cholesky decomposition (for computing the determinant of $1-\chi^0v$), consumes only a small fraction of the total computation time. 

\begin{figure}[htbp]
	\centering
	{
        \includegraphics[width=0.6\textwidth]{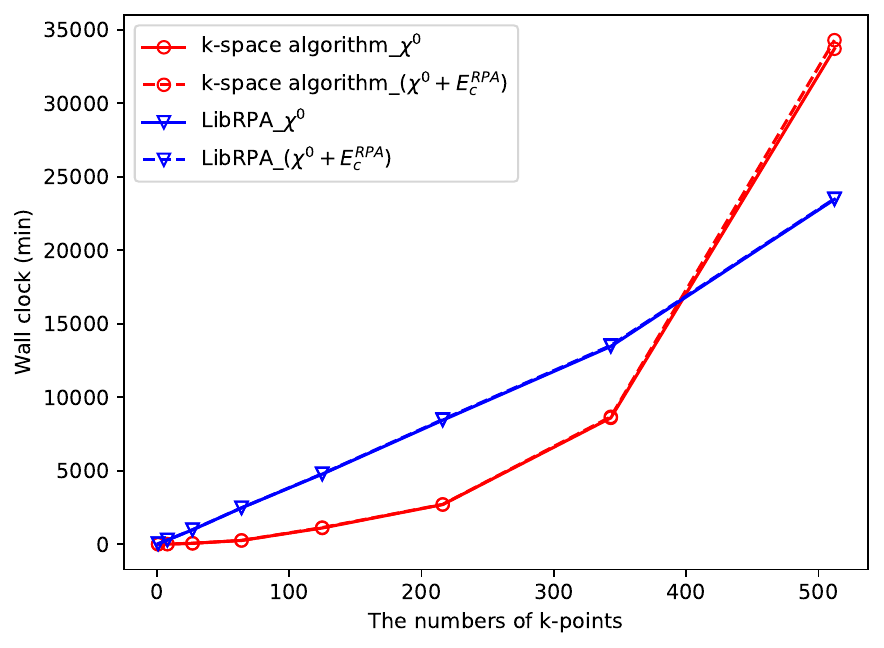} 
    }
      \caption{Scaling behavior of the computation times as a function of number of the $\bfk$ points in the BZ sampling 
       for both the real-space low-scaling algorithm as implemented in LibRPA and the conventional $\bfk$-space algorithm
       as implemented in FHI-aims. The test system is C diamond with a fixed conventional unit cell containing 8 atoms.  All other
       computational settings are the same as Fig.~\ref{fig:scaling-atom}.}   
     \label{fig:scaling-kpt}
\end{figure}

In the above test, only a single $\bfk$ point (i.e., the $\Gamma$ point) is used in the BZ sampling. Such a computational 
setting is suitable for describing
systems with large supercells and low symmetries. Next, we check the scaling behavior of the new algorithm with respect to the number of $\bfk$
points in the BZ, with a fixed unit cell size. In Fig.~\ref{fig:scaling-kpt}, the computational times are presented as a function of 
the number of $\bfk$ points, for both the low-scaling algorithm and conventional $\bfk$-space algorithm. The chosen system in this test
calculation is again the C diamond, albeit with a fixed conventional cell (8 C atoms). Figure~\ref{fig:scaling-kpt} shows that the
computational cost of the real-space algorithm scales linearly with the number of $\bfk$ points, whereas the conventional $\bfk$-space algorithm
scales quadratically, as expected. The crossing point occurs in between the $7\times 7 \times 7$ and $8\times 8 \times 8$ $\bfk$ meshes.  
This suggests that the low-scaling algorithm has an advantage only when very dense $\bfk$ grid is needed. For practical periodic RPA calculations
for simple solids with small unit cells, the conventional $\bfk$-space algorithm is still the preferred method of choice.

\subsection{\label{sec:gf_screening}The Green's-function-based screening}
In the benchmark tests of the scaling behavior presented in Sec.~\ref{sec:scaling_test}, the sparsity of the Green's function matrix 
$G_{ij}(\bfR,\ii\tau)$ is not considered. As such, the low-scaling algorithm in theory scales quadratically 
with respect to system size and linearly with respect to the number of $\bfk$ points. If the sparsity of the Green's function is
further taken into account, as discussed in Algorithm~\ref{alg:algorithm2}, one should achieve an asymptotically linear-scaling behavior
with respect to the system size. In this section, we check how much error in the RPA correlation energy may be incurred if a thresholding parameter
of the Green's-function matrix elements is introduced. From this investigation, one may be able to identify a safe parameter value that
can be used in practical calculations, and find out what additional speedup one can gain if the Green's-function-based
screening is invoked.

\begin{figure}[H]
    \centering
    \includegraphics[width=0.6\textwidth]{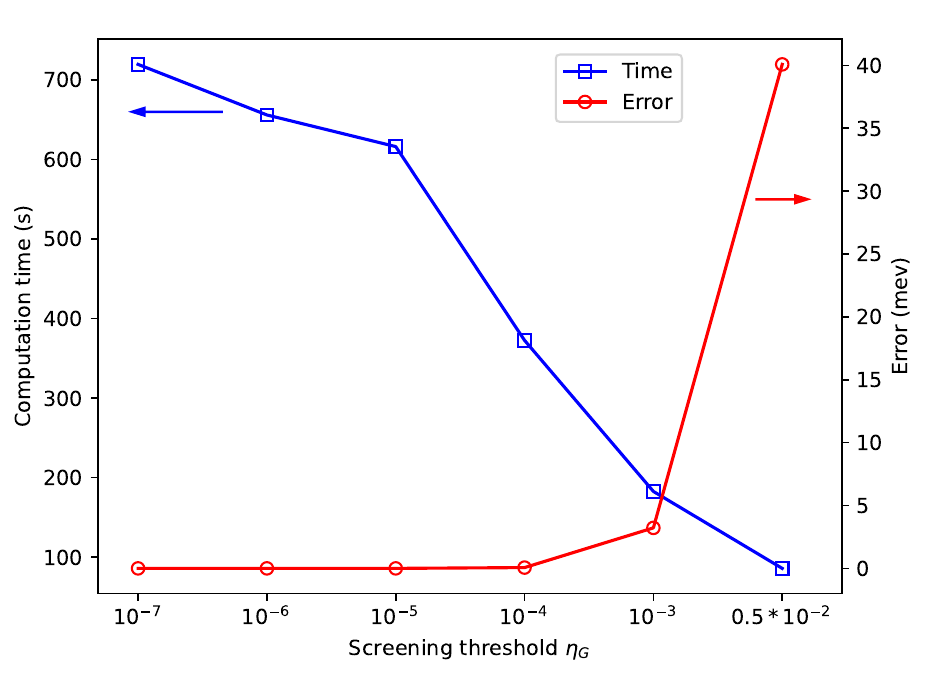}
    \caption{Computation times and errors as a function of the threshold of the Green's function screening. The test system is
    Ar crystal with $6\times 6 \times 6$ $\bfk$ points.
    }
    \label{fig:Green_threshold_Ar}
\end{figure}

Figure~\ref{fig:Green_threshold_Ar} shows the computational time (left $y$ axis) and the error in the computed RPA correlation energy (right $y$ axis) 
as a function of the Green's-function screening parameter $\eta_G$ (introduced in Algorithm~\ref{alg:algorithm2}).
The test system
chosen here is the Ar crystal with $6\times 6 \times 6$ $\bfk$ point mesh (corresponding to a $6\times 6\times 6$ BvK supercell). From Fig.~\ref{fig:Green_threshold_Ar}, one can see that, for such a simple system, a screening parameter of $10^{-4}$ to $10^{-3}$
can lead to one order magnitude reduction of the computational time, yet the incurred error is kept at  meV/atom level for the actual RPA 
calculations. We thus anticipate that the refined low-scaling algorithm that incorporates Green's-function screening will bring significant
additional savings, in particular for wide-gap insulators where the Green's function is expected to quickly decay in real space.

In Appendix~\ref{sec:appendix:GF}, the decay behaviors of the Green's-function matrix elements $G_{ij}(\bfR,\ii\tau)$ for two selected systems -- the Ar crystal and the C diamond crystal are presented.  Figure~\ref{fig:GF_maxV_tau} shows that the Green's function matrix
elements decay rather fast as a function of the distance $d=|\bfR +{\bm \tau}_J - {\bm \tau}_I|$ between the atomic centers. 
The largest amplitude occurs at time $\tau =0$, corresponding
to the density matrix of the system. For Ar, max$\{G_{ij}(\bfR,\ii\tau=0)\}$ becomes vanishingly small for 
$d\ge 10 $ \AA; for C diamond, the decaying of max$\{G_{ij}(\bfR,\ii\tau=0)\}$ is less fast, but its magnitude also becomes
rather small for $d>20$ \AA. 

Moreover, for finite $\tau$, the amplitude of the imaginary-time Green's function decays rather fast as $\tau$ increases,
and becomes tiny for the entire distance range when $\tau>$ 10 a.u, as can be seen from Figs.~\ref{fig:GF_maxV_tau}. 
Thus, the overall decay behavior of the Green's function in real space is governed by the $\tau =0$ case, i.e., the density 
matrix. Consequently, we can expect a linear-scaling behavior of the construction of the $\chi^0$ matrix, within an atomic orbital representation
and LRI approximation. The situation is rather similar to the linear-scaling algorithm developed for the construction of the Hartree-Fock
exchange matrix in terms of NAO basis sets \cite{Levchenko/etal:2015,Lin/Ren/He:2021}. This above line of reasoning applies perfectly to insulating
systems, where the density matrix, and more generally the imaginary-time Green's function, is warranted to decay exponentially in real space. For
metallic systems, the situation is more complicated since the density matrix (and Green's function) decays much slower in real space.
We expect that a linear-scaling behavior can eventually be achieved, but may occur only at very large systems, not in the regime of $10^3$ atoms
that are tested in the present work.

\subsection{Interface with ABACUS}
\begin{figure}[htbp]
    \centering
    \subfigure[]
    {
        \includegraphics[width=0.47\textwidth]{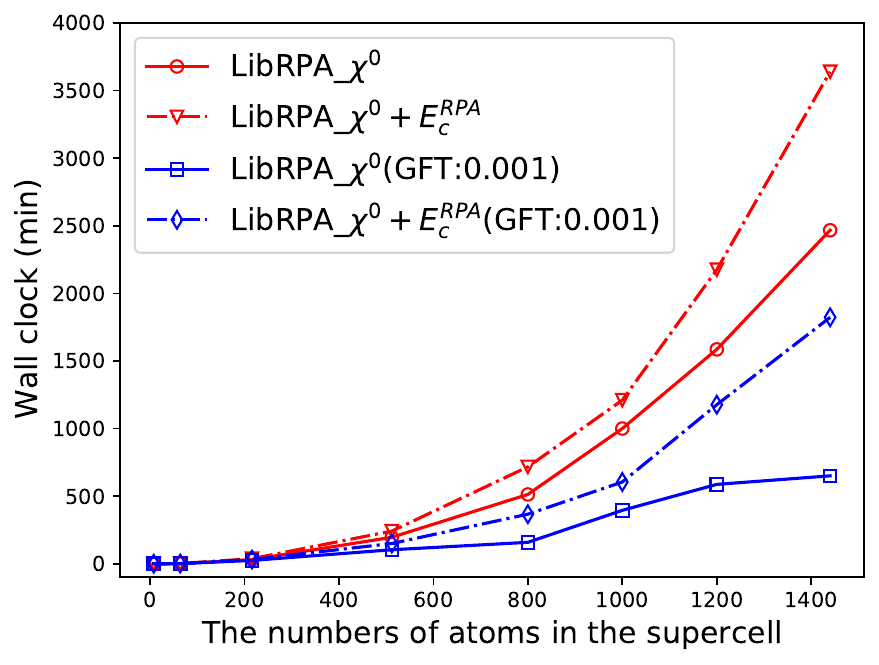}
        \label{fig:abacus_atoms_scaling-a}
    }
    \subfigure[]
    {
        \includegraphics[width=0.47\textwidth]{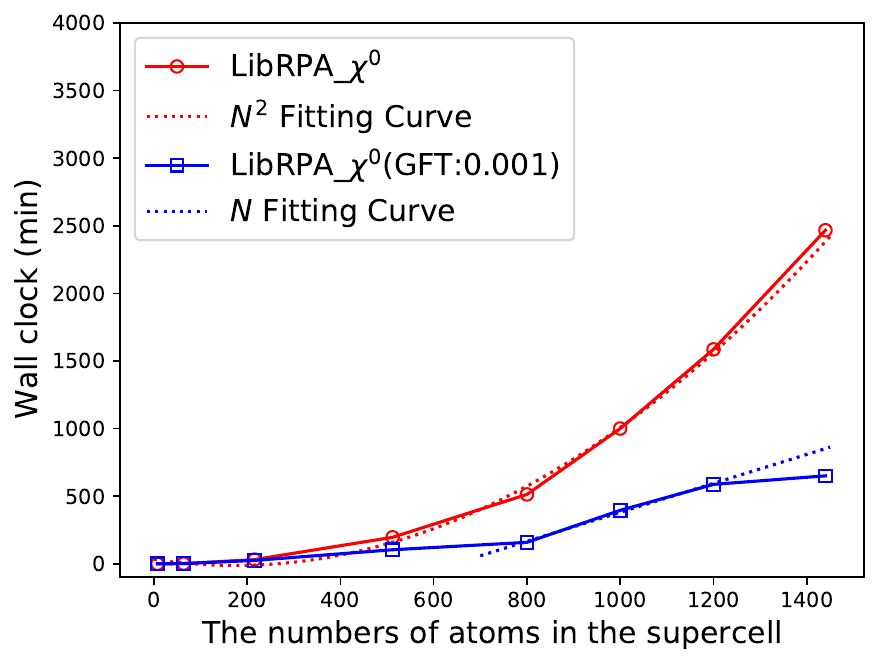}
        \label{fig:abacus_atoms_scaling-b}
    }
    \caption{Scaling behavior the computation times with respect to system size for LibRPA interfaced with ABACUS. 
    The test systems are Si diamond with increasing supercell size. A single $\bfk$ point and the NAO DZP basis set are 
    used is the calculations. Left panel: the timings for both evaluating $\chi^0$ matrix (solid lines) and the total RPA calculation $\chi^0+E_\text{c}^\text{RPA}$ (dash-dotted curves) with (blue curves) and without (red curves) switching on the Green's-function-based screening. 
     Right panel: $\mathcal{O}(N)$- and $\mathcal{O}(N^2)$-scaling curves (dotted lines) are added by fitting the data 
       of the low-scaling algorithms for evaluating the $\chi^0$ matrix with and without Green's-function-based screening,  respectively.}
    \label{fig:abacus_atoms_scaling}
\end{figure}

As a standalone library, LibRPA can also be interfaced with other NAO-based DFT codes besides FHI-aims, provided that
the LRI infrastructure is available.  ABACUS \cite{Chen/Guo/He:2010,Li/Liu/etal:2016} is a DFT software that employs NAOs as its primary basis set choice 
and norm-conserving
pseudopotentials for describing core-valence interactions. In particular, the LRI has been implemented in ABACUS, which enabled efficient
hybrid functional calculations \cite{Lin/Ren/He:2020,Lin/Ren/He:2021,Ji/etal:2022}. As indicated in Fig.~\ref{fig:flowchart}, once the LRI
expansion coefficients $C_{i(\bfo),k(\bfR)}^{\mu(\bfo)}$ 
and the Coulomb matrix $V_{\mu\nu}(\bfk)$  are available, interfacing an NAO-based DFT code with LibRPA is straightforward. 
Figure~\ref{fig:abacus_atoms_scaling} demonstrates the scaling behavior of the computation time of LibRPA interfaced with ABACUS with respect to
system size.
The test systems consist of Si diamond structures of increasing supercell sizes and only a single $\bfk$ point is used. 
The double-$\zeta$ plus polarization (DZP) NAO basis set ($2s2p1d$ for Si) is used in the calculations, whereby the compact basis size allows us 
to go to system size of over 1000 atoms in the supercell. We demonstrate both the computation 
time for evaluating $\chi^0$ matrix and the total time for the RPA correlation energy calculation, with and without turning
on the Green's-function based screening. Two observations are noteworthy: First, the Green's-function based screening starts to have an effect for system
sizes larger than 200 atoms, and can significantly reduce the computational cost for evaluating $\chi^0$; in fact, for system size between
800 and 1400 atoms, the computational cost indeed shows a linear, or even sub-linear scaling with system size, when the Green's-function based screening
is invoked. Second, for system size larger than 800 atoms, the computation of the RPA correlation energy after obtaining the $\chi^0$ matrix, which involves $\mathcal{O}(N^3)$ steps, starts to play a significant role and will eventually dominate the calculations for even larger systems. Thus, for very large systems,
one will also need to develop more efficient lower-scaling algorithms for executing $\chi^0 V$, and for computing the determinant of $1-\chi^0V$. However, this goes beyond the scope of the present paper and will be pursued in future work. 

\section{Summary}
The application of the RPA method to complex materials has been hampered by its quickly increasing computational cost. The rate-determining step for RPA correlation energy calculations in conventional algorithm is the evaluation of the response function matrix $\chi^0$. In this work, 
we present a low-scaling algorithm for evaluating the $\chi^0$ matrix, by combining the real-space, imaginary-time representation of $\chi^0$, 
the strict locality of NAO basis functions, as well as the localized resolution of identity. The algorithm has a formal
$\mathcal{O}(N^2)$ scaling by only taking into account of the sparsity of the LRI expansion coefficients, and becomes linear if the decay behavior
of the Green's function in real space
is further utilized. Benchmark calculations for systems of increasing sizes confirmed the scaling behavior of the proposed algorithm, and benchmark against the conventional $\bfk$-space algorithm confirms the validity and high numerical precision of the present algorithm. 
We particularly show that 
the Green's-function based screening, which has been so far largely overlooked, can bring significant additional
savings for system sizes of over a few hundred atoms.
We also observe that the $\mathcal{O}(N^3)$-scaling steps in RPA calculations after the $\chi^0$ matrix is obtained, whose computational cost is 
negligible in conventional algorithm, starts to dominate for system sizes over 1000 atoms.  Further work is needed to develop low-scaling algorithms
for the $\mathcal{O}(N^3)$ steps. Our work sets a new standard for large-scale periodic RPA 
calculations using atomic orbitals. The low-scaling algorithm we developed and the insights we gained in
the present work not only pushes the limit for
RPA calculations, but are also helpful for extending the reach of other correlated methods to unprecedented size of periodic systems. 



\appendix
\section{Space-time RPA within NAO}
\label{app:equation_derivation}

Equations~(\ref{eq:chi0_matrix}-\ref{eq:chi0_NX_intermidate}) in the main text are the key equations behind our low-scaling RPA algorithm
designed for local atomic basis set framework. These equations are presented in Sec.~\ref{sec:theory} without derivation. Due to their importance
for the entire algorithm, we derive these equations here for completeness.
The starting points are Eqs.~\eqref{eq:chi0_NAOs} and \eqref{eq:LRI} in the main text. Plugging Eq.~\eqref{eq:LRI} into Eq.~\eqref{eq:chi0_NAOs}, 
one obtains
	  \begin{align}
	  \chi^0(\bfr,\bfrp,\ii\tau) = &  \nonumber \\ -\ii\sum_{i,j,k,l}\sum_{\bfR_1,\bfR_2,\bfR_3,\bfR_4} 
		     & \left( \sum_{\mu \in I} C_{i(\bfo),k(\bfR_3-\bfR_1)}^{\mu(\bfo)} P_\mu(\bfr-\bfR_1 - {\bm \tau}_I) + 
		       \sum_{\mu \in K} C_{i(\bfR_1-\bfR_3),k(\bfo)}^{\mu(\bfo)} P_\mu(\bfr-\bfR_3 - {\bm \tau}_K) \right)  \nonumber \\
		    & G_{i,j}(\bfR_2-\bfR_1,\ii\tau) G_{l,k}(\bfR_3-\bfR_4,-\ii\tau) \nonumber \\
    		    &    \left( \sum_{\nu \in J} C_{j(\bfo),l(\bfR_4-\bfR_2)}^{\nu(\bfo)} P_\nu(\bfrp-\bfR_2 -{\bm \tau}_J) + \sum_{\nu \in L} C_{j(\bfR_2-\bfR_4),l(\bfo)}^{\nu(\bfo)}
					   P_\nu(\bfrp-\bfR_4 -{\bm \tau}_L) \right) \nonumber \\
		   & = \chi^{0(A)}(\bfr,\bfrp,\ii\tau)  + \chi^{0(B)}(\bfr,\bfrp,\ii\tau) +  \chi^{0(C)}(\bfr,\bfrp,\ii\tau) +  \chi^{0(D)}(\bfr,\bfrp,\ii\tau) \, ,
                  \label{eq:chi0_decomp}
	  \end{align}
   where the full response function naturally splits into four terms, arising from the special structure due to LRI. These four terms correspond to
   four different ways of placing the ABFs on the four atoms $I$, $J$, $K$, and $L$, within the LRI approximation.
Below we discuss these four terms separately.
In Eq.~\eqref{eq:chi0_decomp}, the first term deals with the situation in which the ABF $\mu$, $\nu$ sit on the atom $I$, $J$ (denoted as
$\mu \in I$ and $\nu \in J$), respectively. 
This term is given by 
	  \begin{align}
		  \chi^{0(A)}(\bfr,\bfrp,i\tau)
		     =& -\ii\sum_{i,j,k,l}\sum_{\bfR_1,\bfR_2,\bfR_3,\bfR_4} \sum_{\mu \in I} C_{i(\bfo),k(\bfR_3-\bfR_1)}^{\mu(\bfo)}
		    P_\mu(\bfr-\bfR_1-{\bm \tau}_I) G_{i,j}(\bfR_2-\bfR_1,\ii\tau) \times \nonumber \\ 
		       & \ \ \ \ \ \ \ \ \ \ \ \ \ \ G_{l,k}(\bfR_3-\bfR_4,-\ii\tau) \sum_{\nu \in J} C_{j(\bfo),l(\bfR_4-\bfR_2)}^{\nu(\bfo)}P_\nu(\bfrp-\bfR_2-{\bm \tau}_J) \nonumber \\
		     =& -\ii \sum_{\mu,\nu,\bfR_1,\bfR_2} P_\mu(\bfr-\bfR_1-{\bm \tau}_I)\sum_{i \in {\cal U}, j \in {\cal V}} \sum_{k,\bfR_3} C_{i(\bfo),k(\bfR_3-\bfR_1)}^{\mu(\bfo)}
			G_{i,j}(\bfR_2-\bfR_1,\ii\tau) \times \nonumber \\
		       & \ \ \ \ \ \ \ \ \ \ \ 	
		        \sum_{l,\bfR_4} \left[ C_{j(\bfo),l(\bfR_4-\bfR_2)}^{\nu(\bfo)} G_{l,k}(\bfR_3-\bfR_4,-\ii\tau) \right] P_\nu(\bfrp-\bfR_2-{\bm \tau}_J) \nonumber \\
		     = & \sum_{\mu,\nu,\bfR_1,\bfR_2} P_\mu(\bfr-\bfR_1-{\bm \tau}_{\cal U}) \chi^{0(A)}_{\mu,\nu} (\bfR_2-\bfR_1,\ii\tau) P_\nu(\bfrp-\bfR_2-
       {\bm \tau}_{\cal V}) 
	  \end{align}
where
      \begin{equation}
	      \chi^{0(A)}_{\mu,\nu} (\bfR_2-\bfR_1,\ii\tau) =-\ii \sum_{i \in {\cal U}, j \in {\cal V}} \sum_{k,\bfR_3} \sum_{l,\bfR_4} C_{i(\bfo),k(\bfR_3-\bfR_1)}^{\mu(\bfo)}
				       G_{i,j}(\bfR_2-\bfR_1,\ii\tau) C_{j(\bfo),l(\bfR_4-\bfR_2)}^{\nu(\bfo)} G_{l,k}(\bfR_3-\bfR_4,-\ii\tau) \, .
      \end{equation}
Recall that ${\cal U}$ and ${\cal V}$ denote the atoms where the ABFs $\mu$, $\nu$ are centering, and ${\bm \tau}_{\cal U}$ and ${\bm \tau}_{\cal V}$ are their respective atomic positions in the unit cell. In the above derivation, we have used the fact that, in the present situation,
the atom ${\cal U}=I$, and ${\cal V}=J$ (and hence ${\bm \tau}_{\cal U}={\bm \tau}_I$, and ${\bm \tau}_{\cal U}={\bm \tau}_J$). We have also
used the property that, in the computation of $\chi^{0(A)}(\bfr,\bfrp,i\tau)$,  first looping over the AOs $i,j$ and requiring $\mu \in I$ and $\nu \in J$ is equivalent to first looping over the ABFs 
$\mu$, $\nu$, and requiring the AOs $i \in {\cal U}$ and $j \in {\cal V}$.
      Making use of the translational symmetry, we can, without losing generality, set $\bfR =\bfR_2 - \bfR_1$ and $\bfR_1 = \bfo$. Finally
      we obtain
      \begin{equation}
	      \chi^{0(A)}_{\mu,\nu} (\bfR,\ii\tau) =-\ii \sum_{i \in {\cal U}, j \in {\cal V}} \sum_{k,\bfR_3} \sum_{l,\bfR_4} C_{i(\bfo),k(\bfR_3)}^{\mu(\bfo)}
					   G_{l,k}(\bfR_3-\bfR_4,-\ii\tau) C_{j(\bfo),l(\bfR_4-\bfR)}^{\nu(\bfo)} G_{i,j}(\bfR,\ii\tau) \, .
	 \label{eq:chi_0_A}
      \end{equation}

Next, we deal with the second term that corresponds to the situation in which $\mu \in I$ and $\nu \in L$ (i.e., $I={\cal U}$ and $L={\cal V}$).
Specifically,
	  \begin{align}
		  \chi^{0(B)}(\bfr,\bfrp,\ii\tau)
		     =& -\ii\sum_{i,j,k,l}\sum_{\bfR_1,\bfR_2,\bfR_3,\bfR_4} \sum_{\mu \in I} C_{i(\bfo),k(\bfR_3-\bfR_1)}^{\mu(\bfo)}
		    P_\mu(\bfr-\bfR_1 - {\bm \tau}_I) G_{i,j}(\bfR_2-\bfR_1,\ii\tau) \times  \nonumber \\ 
		       & \ \ \ \ \ \ \ \ \ \ \ \ \ G_{l,k}(\bfR_3-\bfR_4,-\ii\tau)  \sum_{\nu \in L} C_{j(\bfR_2-\bfR_4),l(\bfo)}^{\nu(\bfo)}P_\nu(\bfrp-\bfR_4 - {\bm \tau}_L) \nonumber \\
		    \overset{\bfR_2 \leftrightarrow \bfR_4}{=}& -\ii\sum_{i,j,k,l}\sum_{\bfR_1,\bfR_2,\bfR_3,\bfR_4} \sum_{\mu \in I} C_{i(\bfo),k(\bfR_3-\bfR)_1}^{\mu(\bfo)}
		    P_\mu(\bfr-\bfR_1 - {\bm \tau}_I) G_{i,j}(\bfR_4-\bfR_1,\ii\tau) \times \nonumber \\  
		       & \ \ \ \ \ \ \ \ \ \ \ \ \ G_{l,k}(\bfR_3-\bfR_2,-\ii\tau) \sum_{\nu \in L} C_{j(\bfR_4-\bfR_2),l(\bfo)}^{\nu(\bfo)}P_\nu(\bfrp-\bfR_2 - {\bm \tau}_L ) \ \ \ \ 
		        \nonumber \\
		     =& -\ii \sum_{\mu,\nu,\bfR_1,\bfR_2 } P_\mu(\bfr-\bfR_1 -{\bm \tau}_{\cal U}  )\sum_{i \in {\cal U}, l \in {\cal V}} \sum_{k,\bfR_3} C_{i(\bfo),k(\bfR_3-\bfR_1)}^{\mu(\bfo)}
			G_{i,j}(\bfR_4-\bfR_1,\ii\tau) \times \nonumber \\
		       & \ \ \ \ \ \ \ \ \ \ \ 	
		        \sum_{j,\bfR_4} C_{j(\bfR_4-\bfR_2),l(\bfo)}^{\nu(\bfo)} G_{l,k}(\bfR_3-\bfR_2,-\ii\tau) P_\nu(\bfrp-\bfR_2-{\bm \tau}_{\cal V} ) \nonumber \\
		     = & \sum_{\mu,\nu,\bfR_1,\bfR_2} P_\mu(\bfr-\bfR_1 -{\bm \tau}_{\cal U}) \chi^{0(B)}_{\mu,\nu} (\bfR_2-\bfR_1,\ii\tau) 
                   P_\nu(\bfrp-\bfR_2-{\bm \tau}_{\cal V}) 
                \label{eq:app-termB}
	  \end{align}
where
      \begin{equation}
	      \chi^{0(B)}_{\mu,\nu} (\bfR_2-\bfR_1,\ii\tau) =-\ii \sum_{i \in {\cal U}, l \in {\cal V}} \sum_{k,\bfR_3} \sum_{j,\bfR_4} C_{i(\bfo),k(\bfR_3-\bfR_1)}^{\mu(\bfo)}G_{i,j}(\bfR_4-\bfR_1,\ii\tau)C_{j(\bfR_4-\bfR_2),l(\bfo)}^{\nu(\bfo)} G_{l,k}(\bfR_3-\bfR_2,-\ii\tau)
	 \end{equation}
      or
      \begin{equation}
	      \chi^{0(B)}_{\mu,\nu} (\bfR,i\tau) =-\ii \sum_{i \in {\cal U}, l \in {\cal V}} \sum_{k,\bfR_3} \sum_{j,\bfR_4} C_{i(\bfo),k(\bfR_3)}^{\mu(\bfo)}
					   G_{l,k}(\bfR_3-\bfR,-\ii\tau) C_{j(\bfR_4-\bfR),l(\bfo)}^{\nu(\bfo)} G_{i,j}(\bfR_4,\ii\tau) \, .	  
	 \label{eq:chi_0_B}
      \end{equation}
      In the derivation of Eq.~\eqref{eq:app-termB}, we again used the property that, in the present case, looping over the AOs $i,l$ and requiring 
      $\mu \in I$ and $\nu \in L$ is equivalent to first looping over the ABFs $\mu,\nu$ and requiring $i \in {\cal U}$ and $l \in {\cal V}$.

      Next comes the third term, corresponding to the situation where $\mu \in K$ and $\nu \in J$ (i.e., $K={\cal U}$ and $J={\cal V}$),
	  \begin{align}
		  \chi^{0(C)}(\bfr,\bfrp,\ii\tau)
		     =& -\ii\sum_{i,j,k,l}\sum_{\bfR_1,\bfR_2,\bfR_3,\bfR_4} \sum_{\mu \in K} C_{i(\bfR_1-\bfR_3),k(\bfo)}^{\mu(\bfo)}
		    P_\mu(\bfr-\bfR_3 -{\bm \tau}_K) G_{i,j}(\bfR_2-\bfR_1,\ii\tau) \times \nonumber \\  
		       & \ \ \ \ \ \ \ \ \ \ \ \  G_{l,k}(\bfR_3-\bfR_4,-\ii\tau) \sum_{\nu \in J} 
		       C_{j(\bfo),l(\bfR_4-\bfR_2)}^{\nu(\bfo)}P_\nu(\bfrp-\bfR_2 - {\bm \tau}_J) \nonumber \\
		    \overset{\bfR_1\leftrightarrow \bfR_3}{=}& -\ii\sum_{i,j,k,l}\sum_{\bfR_1,\bfR_2,\bfR_3,\bfR_4} \sum_{\mu \in K} C_{i(\bfR_3-\bfR_1),k(\bfo)}^{\mu(\bfo)}
		    P_\mu(\bfr-\bfR_1 -{\bm \tau}_K ) G_{i,j}(\bfR_2-\bfR_3,\ii\tau) \times \nonumber \\  
		       & \ \ \ \ \ \ \ \ \ \ \ \ G_{l,k}(\bfR_1-\bfR_4,-\ii\tau) \sum_{\nu \in J} C_{j(\bfo),l(\bfR_4-\bfR_2)}^{\nu(\bfo)}P_\nu(\bfrp-\bfR_2 -{\bm \tau}_J) 
		        \nonumber \\
		     =& -\ii \sum_{\mu,\nu,\bfR_1,\bfR_2} P_\mu(\bfr-\bfR_1 -{\bm \tau}_{\cal U} )\sum_{k \in {\cal U}, j \in {\cal V}} \sum_{i,\bfR_3} \sum_{l,\bfR_2} C_{i(\bfR_3-\bfR_1),k(\bfo)}^{\mu(\bfo)}
			G_{i,j}(\bfR_2-\bfR_3,\ii\tau) \times \nonumber \\
		       & \ \ \ \ \ \ \ \ \ \ \ \ \ \ \ 
		         C_{j(\bfo),l(\bfR_4-\bfR_2)}^{\nu(\bfo)} G_{l,k}(\bfR_1-\bfR_4,-\ii\tau) P_\nu(\bfrp-\bfR_2-{\bm \tau}_{\cal V}) \nonumber \\
		     = & \sum_{\mu,\nu,\bfR_1,\bfR_2} P_\mu(\bfr-\bfR_1-{\bm \tau}_{\cal U}) \chi^{0(C)}_{\mu,\nu} (\bfR_2-\bfR_1,\ii\tau) P_\nu(\bfrp-\bfR_2-{\bm \tau}_{\cal V} ) \, ,
	  \end{align}
where
      \begin{equation}
	      \chi^{0(C)}_{\mu,\nu} (\bfR_2-\bfR_1,\ii\tau) = -\ii\sum_{k \in {\cal U}, j \in {\cal V}} \sum_{i,\bfR_3} \sum_{l,\bfR_4} C_{i(\bfR_3-\bfR_1),k(\bfo)}^{\mu(\bfo)} G_{i,j}(\bfR_2-\bfR_3,\ii\tau)
	     C_{j(\bfo),l(\bfR_4-\bfR_2)}^{\nu(\bfo)} G_{l,k}(\bfR_1-\bfR_4,-\ii\tau) \, ,
	 \end{equation}
      or 
      \begin{equation}
	     \chi^{0(C)}_{\mu,\nu} (\bfR,i\tau) = -\ii\sum_{k \in {\cal U}, j \in {\cal V}} \sum_{i,\bfR_3} \sum_{l,\bfR_4} C_{i(\bfR_3),k(\bfo)}^{\mu(\bfo)} G_{i,j}(\bfR-\bfR_3,\ii\tau)
	     C_{j(\bfo),l(\bfR_4-\bfR)}^{\nu(\bfo)} G_{l,k}(-\bfR_4,-\ii\tau) \, .
	     \label{eq:chi_0_C}
      \end{equation}     
      
      Finally we deal with the fourth term, which corresponds to the situation where $\mu \in K$ and $\nu \in L$ (i.e., $K={\cal U}$ and $L={\cal V}$),
	  \begin{align}
		  \chi^{0(D)}(\bfr,\bfrp,\ii\tau)
		     =& -\ii\sum_{i,j,k,l}\sum_{\bfR_1,\bfR_2,\bfR_3,\bfR_4} \sum_{\mu \in K} C_{i(\bfR_1-\bfR_3),k(\bfo)}^{\mu(\bfo)}
		    P_\mu(\bfr-\bfR_3 -{\bm \tau}_K) G_{i,j}(\bfR_2-\bfR_1,\ii\tau) \times \nonumber \\  
		       & \ \ \ \ \ \ \ \ \ \ \ \ \  G_{l,k}(\bfR_3-\bfR_4,-\ii\tau) \sum_{\nu \in L} C_{j(\bfR_2-\bfR_4),l(\bfo)}^{\nu(\bfo)}P_\nu(\bfrp-\bfR_4-{\bm \tau}_K) \nonumber \\
		    \overset{\bfR_1\leftrightarrow \bfR_3, \bfR_2 \leftrightarrow \bfR_4}{=}& -\ii\sum_{i,j,k,l}\sum_{\bfR_1,\bfR_2,\bfR_3,\bfR_4} \sum_{\mu \in K} C_{i(\bfR_3-\bfR_1),k(\bfo)}^{\mu(\bfo)}
		    P_\mu(\bfr-\bfR_1-{\bm \tau}_K ) G_{i,j}(\bfR_4-\bfR_3,\ii\tau) \times \nonumber \\  
		       & \ \ \ \ \ \ \ \ \ \ \ \ \  G_{l,k}(\bfR_1-\bfR_2,-\ii\tau) \sum_{\nu \in L} C_{j(\bfR_4-\bfR_2),l(\bfo)}^{\nu(\bfo)}P_\nu(\bfrp-\bfR_2-{\bm \tau}_L ) 
		        \nonumber \\
		     =& -\ii \sum_{\mu,\nu,\bfR_1,\bfR_2} P_\mu(\bfr-\bfR_1 - {\bm \tau}_{\cal U})\sum_{k \in {\cal U}, l \in {\cal V}} \sum_{i,\bfR_3} \sum_{j,\bfR_4} C_{i(\bfR_3-\bfR_1),k(\bfo)}^{\mu(\bfo)}
			G_{i,j}(\bfR_4-\bfR_3,\ii\tau) \times \nonumber \\
		       & \ \ \ \ \ \ \ \ \  \ \ \ \
		         C_{j(\bfR_4-\bfR_2),l(\bfo)}^{\nu(\bfo)} G_{l,k}(\bfR_1-\bfR_2,-\ii\tau) P_\nu(\bfrp-\bfR_2 -{\bm \tau}_{\cal U} ) \nonumber \\
		     = & \sum_{\mu,\nu,\bfR_1,\bfR_2} P_\mu(\bfr-\bfR_1- {\bm \tau}_{\cal U}) 
                  \chi^{0(D)}_{\mu,\nu} (\bfR_2-\bfR_1,\ii\tau) P_\nu(\bfrp-\bfR_2 - {\bm \tau}_{\cal V} ) \, ,  
	  \end{align}
where
      \begin{equation}
	      \chi^{0(D)}_{\mu,\nu} (\bfR_2-\bfR_1,i\tau) =-\ii \sum_{k \in {\cal U}, l \in {\cal V}} \sum_{i,\bfR_3} \sum_{j,\bfR_4} C_{i(\bfR_3-\bfR_1),k(\bfo)}^{\mu(\bfo)} G_{i,j}(\bfR_4-\bfR_3,\ii\tau)
	     C_{j(\bfR_4-\bfR_2),l(\bfo)}^{\nu(\bfo)} G_{l,k}(\bfR_1-\bfR_2,-\ii\tau) \, , 
	 \end{equation}
      or 
      \begin{equation}
	     \chi^{0(D)}_{\mu,\nu} (\bfR,i\tau) =-\ii \sum_{k \in {\cal U}, l \in {\cal V}} \sum_{i,\bfR_3} \sum_{j,\bfR_4} C_{i(\bfR_3),k(\bfo)}^{\mu(\bfo)} G_{i,j}(\bfR_4-\bfR_3,\ii\tau)
	     C_{j(\bfR_4-\bfR),l(\bfo)}^{\nu(\bfo)} G_{l,k}(-\bfR,-\ii\tau)
	     \label{eq:chi_0_D}  \, .
      \end{equation}     

Summing up Eqs.~\eqref{eq:chi_0_A}, \eqref{eq:chi_0_B}, \eqref{eq:chi_0_C}, and \eqref{eq:chi_0_D}, we obtain,
 \begin{align}
 	      \chi^{0}_{\mu,\nu} (\bfR,i\tau) & = \chi^{0(A)}_{\mu,\nu} (\bfR,\ii\tau)+
 	      \chi^{0(B)}_{\mu,\nu} (\bfR,\ii\tau)+
 	      \chi^{0(C)}_{\mu,\nu} (\bfR,\ii\tau)+\chi^{0(D)}_{\mu,\nu} (\bfR,\ii\tau)  \nonumber \\
 	    = & -\ii  \left[ \sum_{i \in {\cal U}, j \in {\cal V}} \sum_{k,\bfR_3} \sum_{l,\bfR_4} 
 	              C_{i(\bfo),k(\bfR_3)}^{\mu(\bfo)}
			   G_{l,k}(\bfR_3-\bfR_4,-\ii\tau) C_{j(\bfo),l(\bfR_4-\bfR)}^{\nu(\bfo)} G_{i,j}(\bfR,\ii\tau)
			    \right. \nonumber \\ 
 	     & + \sum_{i \in {\cal U}, l \in {\cal V}} \sum_{k,\bfR_3} \sum_{j,\bfR_4} 
 	    C_{i(\bfo),k(\bfR_3)}^{\mu(\bfo)}
		 G_{l,k}(\bfR_3-\bfR,-\ii\tau) C_{j(\bfR_4-\bfR),l(\bfo)}^{\nu(\bfo)} G_{i,j}(\bfR_4,\ii\tau) 
		 \nonumber \\
  	    &  +\sum_{k \in {\cal U}, j \in {\cal V}} \sum_{i,\bfR_3} \sum_{l,\bfR_4} C_{i(\bfR_3),k(\bfo)}^{\mu(\bfo)} G_{i,j}(\bfR-\bfR_3,\ii\tau)
	     C_{j(\bfo),l(\bfR_2-\bfR)}^{\nu(\bfo)} G_{l,k}(-\bfR_4,-\ii\tau) \nonumber \\
 	     &\left. + \sum_{k \in {\cal U}, l \in {\cal V}} \sum_{i,\bfR_3} \sum_{j,\bfR_4} C_{i(\bfR_3),k(\bfo)}^{\mu(\bfo)} G_{i,j}(\bfR_4-\bfR_3,\ii\tau)
	     C_{j(\bfR_4-\bfR),l(\bfo)}^{\nu(\bfo)} G_{l,k}(-\bfR,-\ii\tau) \right]
 \end{align}
 
To facilitate its computation and in particular the design of the loop structure in the low-scaling algorithm, we swap the dummy indices in the
summation. Specifically, we perform the following exchanges for orbital 
indices: $j\leftrightarrow l$ for the second term, $i\leftrightarrow k$ for the third term,
$i\leftrightarrow k, j\leftrightarrow l$ for the fourth term. And for all terms, further making the following replacement
for the lattice vectors: $\bfR_3 \rightarrow \bfR_1$, and $\bfR_4 \rightarrow \bfR_2$, we have
 \begin{align}
	\chi_{\mu,\nu}^0(\textbf{R},\ii\tau) 
	&= -\ii \left[\sum_{i \in \mathcal{U},j \in\mathcal{V}}\sum_{k,\textbf{R}_1}\sum_{l,\textbf{R}_2}C_{i(\textbf{0}),k(\textbf{R}_1)}^{\mu (\textbf{0})}
	G_{l,k}(\textbf{R}_1-\textbf{R}_2,-\ii\tau)C_{j(\textbf{0}),l(\textbf{R}_2-\textbf{R})}^{\nu (\textbf{0})}G_{i,j}(\textbf{R},\ii\tau)\right. \nonumber\\
	&\quad +\sum_{i \in \mathcal{U},j \in\mathcal{V}}\sum_{k,\textbf{R}_1}\sum_{l,\textbf{R}_2}C_{i(\textbf{0}),k(\textbf{R}_1)}^{\mu (\textbf{0})}
	G_{j,k}(\textbf{R}_1-\textbf{R},-\ii\tau)C_{l(\textbf{R}_2-\textbf{R}),j(\textbf{0})}^{\nu (\textbf{0})}G_{i,l}(\textbf{R}_2,\ii\tau) \nonumber\\
	&\quad +\sum_{i \in \mathcal{U},j \in\mathcal{V}}\sum_{k,\textbf{R}_1}\sum_{l,\textbf{R}_2}C_{k(\textbf{R}_1),i(\textbf{0})}^{\mu (\textbf{0})}
	G_{k,j}(\textbf{R}-\textbf{R}_1,\ii\tau)C_{j(\textbf{0}),l(\textbf{R}_2-\textbf{R})}^{\nu (\textbf{0})}G_{l,i}(-\textbf{R}_2,-\ii\tau) \nonumber \\
	&\left. \quad +\sum_{i \in \mathcal{U},j \in\mathcal{V}}\sum_{k,\textbf{R}_1}\sum_{l,\textbf{R}_2}C_{k(\textbf{R}_1),i(\textbf{0})}^{\mu (\textbf{0})}
	G_{k,l}(\textbf{R}_2-\textbf{R}_1,\ii\tau)C_{l(\textbf{R}_2-\textbf{R}),j(\textbf{0})}^{\nu (\textbf{0})}G_{j,i}(-\textbf{R},-\ii\tau)\right] \nonumber\\
	&= -\ii \left[\sum_{i \in \mathcal{U},j \in\mathcal{V}}\sum_{k,\textbf{R}_1}\sum_{l,\textbf{R}_2}C_{i(\textbf{0}),k(\textbf{R}_1)}^{\mu (\textbf{0})}\left(
	G_{l,k}(\textbf{R}_1-\textbf{R}_2,-\ii\tau)C_{j(\textbf{0}),l(\textbf{R}_2-\textbf{R})}^{\nu (\textbf{0})}G_{i,j}(\textbf{R},\ii\tau)\right. \right.\nonumber\\
	&\qquad\qquad\qquad\qquad\qquad\qquad\quad +G_{j,k}(\textbf{R}_1-\textbf{R},-\ii\tau)C_{l(\textbf{R}_2-\textbf{R}),j(\textbf{0})}^{\nu (\textbf{0})}G_{i,l}(\textbf{R}_2,\ii\tau)\nonumber\\
	&\qquad\qquad\qquad\qquad\qquad\qquad\quad +G^*_{j,k}(\textbf{R}_1-\textbf{R},\ii\tau)C_{j(\textbf{0}),l(\textbf{R}_2-\textbf{R})}^{\nu (\textbf{0})}G_{i,l}(\textbf{R}_2,-\ii\tau)\nonumber\\
	&\left. \left.\qquad\qquad\qquad\qquad\qquad\qquad\quad +G^*_{l,k}(\textbf{R}_1-\textbf{R}_2,\ii\tau)C_{l(\textbf{R}_2-\textbf{R}),j(\textbf{0})}^{\nu (\textbf{0})}G^*_{i,j}(\textbf{R},-\ii\tau)\right) \right] \, ,
 \label{eq:app_chi0_fourterms}
 \end{align}
where we have utilized the symmetry properties for the
Green's function, i.e., $G_{i,j}(\textbf{R},\ii\tau)=G_{j,i}^*(-\textbf{R},\ii\tau)$, and for the expansion coefficients, i.e.,
$C_{i(\textbf{0}),k(\textbf{R}_1)}^{\mu (\textbf{0})}$. Close inspection of the four terms in Eq.~\eqref{eq:app_chi0_fourterms} suggests 
that the first and fourth terms can be grouped together, and so do the second and third terms. Namely, 
 \begin{align}
 	\chi_{\mu,\nu}^0(\textbf{R},\ii\tau) 
     &=-\ii \left[\sum_{i \in \mathcal{U}}\sum_{k,\textbf{R}_1}C_{i(\textbf{0}),k(\textbf{R}_1)}^{\mu (\textbf{0})}\left( \sum_{j \in \mathcal{V}}G_{i,j}(\textbf{R},\ii\tau)  \sum_{l,\textbf{R}_2}C_{j(\textbf{0}),l(\textbf{R}_2-\textbf{R})}^{\nu (\textbf{0})}G_{l,k}(\textbf{R}_1-\textbf{R}_2,-\ii\tau)  \right. \right.\nonumber\\
	&\qquad\qquad\qquad\qquad\qquad +\sum_{j \in \mathcal{V}}G^*_{i,j}(\textbf{R},-\ii\tau)  \sum_{l,\textbf{R}_2}C_{j(\textbf{0}),l(\textbf{R}_2-\textbf{R})}^{\nu (\textbf{0})}G^*_{l,k}(\textbf{R}_1-\textbf{R}_2,\ii\tau)\nonumber\\
	&\qquad\qquad\qquad\qquad\qquad +\sum_{j \in \mathcal{V}}G_{j,k}(\textbf{R}_1-\textbf{R},-\ii\tau)\sum_{l,\textbf{R}_2}C_{j(\textbf{0}),l(\textbf{R}_2-\textbf{R})}^{\nu (\textbf{0})}G_{i,l}(\textbf{R}_2,\ii\tau)\nonumber\\
	&\qquad\qquad\qquad\qquad\qquad\left.\left. +\sum_{j \in \mathcal{V}}G^*_{j,k}(\textbf{R}_1-\textbf{R},\ii\tau)\sum_{l,\textbf{R}_2}C_{j(\textbf{0}),l(\textbf{R}_2-\textbf{R})}^{\nu (\textbf{0})}G^*_{i,l}(\textbf{R}_2,-\ii\tau) \right) \right]\nonumber \\
	&=-\ii \Bigg[\sum_{i \in \mathcal{U}}\sum_{k,\textbf{R}_1}C_{i(\textbf{0}),k(\textbf{R}_1)}^{\mu (\textbf{0})}\left(  M^{\nu}_{i,k}(\textbf{R}_1,\textbf{R},\ii\tau)+ M^{\nu*}_{i,k}(\textbf{R}_1,\textbf{R},-\ii\tau)\right.\nonumber\\
	&\qquad\qquad\qquad\qquad\qquad\left.+ Z^{\nu}_{i,k}(\textbf{R}_1,\textbf{R},\ii\tau)+ Z^{\nu*}_{i,k}(\textbf{R}_1,\textbf{R},-\ii\tau) \right)\Bigg]
\end{align}

where

\begin{eqnarray}
M^{\nu}_{i,k}(\textbf{R}_1,\textbf{R},\ii\tau)&=&\sum_{j \in \mathcal{V}}G_{i,j}(\textbf{R},\ii\tau)N^{\nu}_{j,k}(\textbf{R}_1,\textbf{R},\ii\tau) \nonumber \\
N^{\nu}_{j,k}(\textbf{R}_1,\textbf{R},\ii\tau)&=&\sum_{l,\textbf{R}_2}C_{j(\textbf{0}),l(\textbf{R}_2-\textbf{R})}^{\nu (\textbf{0})}G_{l,k}(\textbf{R}_1-\textbf{R}_2,-\ii\tau) \nonumber
\end{eqnarray}
and 
\begin{eqnarray}
Z^{\nu}_{i,k}(\textbf{R}_1,\textbf{R},\ii\tau)&=&\sum_{j \in \mathcal{V}}G_{j,k}(\textbf{R}_1-\textbf{R},-\ii\tau)X^\nu_{i,j}(\textbf{R},\ii\tau) \nonumber \\
X^\nu_{i,j}(\textbf{R},\ii\tau)&=&\sum_{l,\textbf{R}_2}C_{j(\textbf{0}),l(\textbf{R}_2-\textbf{R})}^{\nu (\textbf{0})}G_{i,l}(\textbf{R}_2,\ii\tau) 
\nonumber 
\end{eqnarray}
Hence, Eqs~(\ref{eq:chi0_matrix}-\ref{eq:chi0_NX_intermidate}) in the main text are derived.

\section{Decay behavior of the Green's function in real space}
\label{sec:appendix:GF}
\begin{figure}[htbp]
	\centering
	{
        \includegraphics[width=\textwidth]{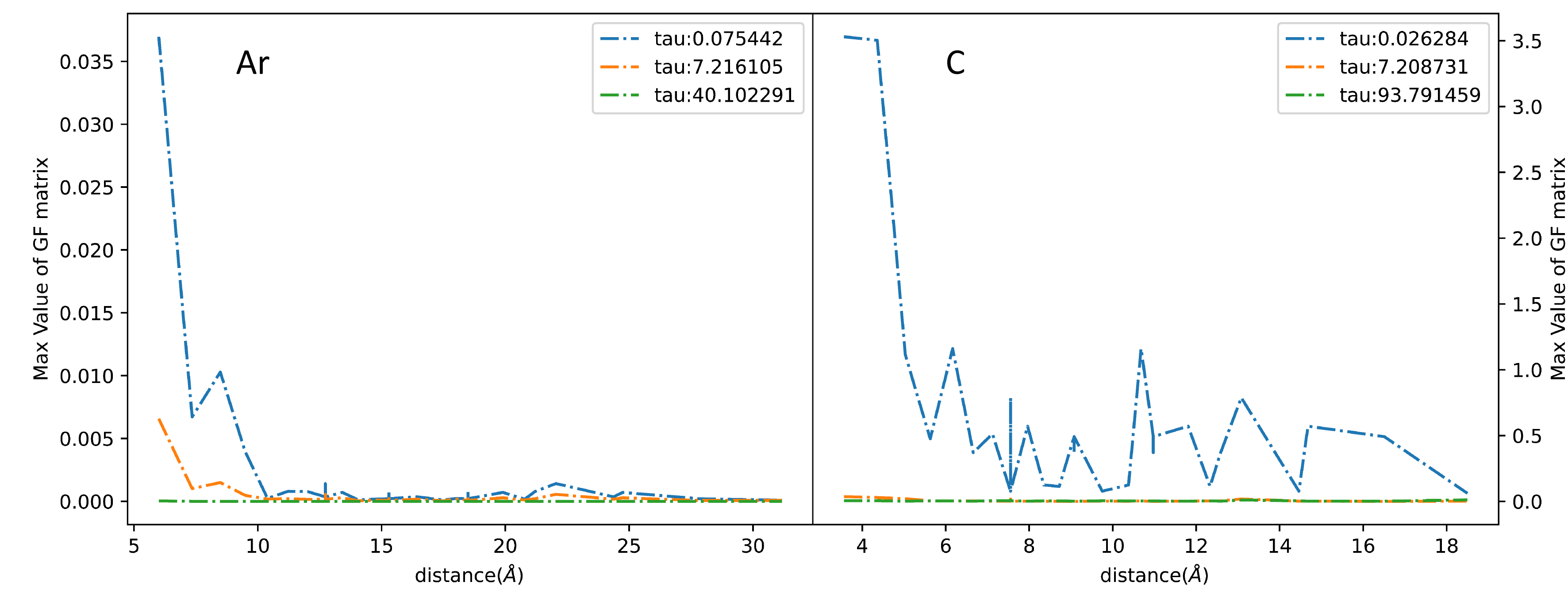} 
    }
      \caption{The maximal matrix elements of Green's function as a function of the distance between the atomic centers for Ar crystal (left panel) and diamond (right panel). The calculations were done with ABACUS using NAO DZP basis set and $6\times6\times6$ $\bfk$ point mesh.}   
     \label{fig:GF_maxV_tau}
\end{figure}

In Sec.~\ref{sec:gf_screening}, we showed that drastic further computational savings can be achieved if Green's-function based screening
is incorporated. The scaling behavior of the refined real-space algorithm depends on the decay behavior of the Green's function in real space.
In Fig.~\ref{fig:GF_maxV_tau}, we present the absolute values of the maximal matrix elements of
the imaginary-time Green's function $G_{ij}(\bfR, \ii\tau)$ as a function of the
distance between the atomic centers for the Ar crystal (left panel) and C diamond crystal (right panel), respectively. Namely, what is plotted are
$\max_{i\in I, j\in J} {|G_{ij}(\bfR, \ii\tau)|}$ as a function of $d=|\bfR + {\bm \tau }_J - {\bm \tau}_I|$ at three 
different time points, i.e., $\tau = 0.075442, 7.216105, 40.102291$ a.u. for Ar and $\tau = 0.026284, 7.208731, 93.791459$ a.u. for C diamond.

\bibliographystyle{apsrev4-2}
\bibliography{CommonBib}
\end{document}

%% file: newcommands.tex
\newcommand{\bfr}{ {\mathbf r}} 
\newcommand{\bfrp}{ {\mathbf r'}} 
\newcommand{\bfzero}{ {\bf 0}} 
\newcommand{\bfR}{ {\mathbf R}} 
\newcommand{\bfo}{ {\bf 0}} 
\newcommand{\bfq}{ {\mathbf q}} 
\newcommand{\bfk}{ {\mathbf k}} 

\newcommand{\ii}{\mathrm{i}}
\newcommand{\ee}{\mathrm{e}}